\documentclass[twocolumn,preprintnumbers,amsmath,amssymb]{revtex4}

\usepackage{graphicx}
\usepackage{dcolumn}
\usepackage{bm}

\begin{document}

\title{An experimental measurement of the coexistence curve\\ and critical temperature, density and pressure of bulk nuclear matter}

\author{J. B. Elliott$^1$, P. T. Lake$^2$, L. G. Moretto$^2$ and L. Phair$^2$}
\affiliation{$^1$Lawrence Livermore National Laboratory, 7000 East Avenue, Livermore, CA 94550\\ $^2$Lawrence Berkeley National Laboratory, 1 Cyclotron Road, Berkeley CA 94720}

\preprint{LLNL-JRNL-539511-DRAFT}

\date{\today}
\begin{abstract}
Infinite, neutron-proton symmetric, neutral nuclear matter has a critical temperature of $T_c = 17.9 \pm 0.4$ MeV, a critical density of $\rho_c =  0.06 \pm 0.01$ A$/$fm$^3$ and a critical pressure of $p_c = 0.31\pm0.07$ MeV$/$fm$^3$.\ \ These values have been obtained from our analysis of data from six different reactions studied in three different experiments: two ``compound nuclear'' reactions: $^{58}$Ni$+^{12}$C$\rightarrow^{70}$Se and $^{64}$Ni$+^{12}$C$\rightarrow^{76}$Se (both performed at the LBNL 88" Cyclotron) and four ``multifragmentation'' reactions: 1 GeV$/$c $\pi+^{197}$Au (performed by the ISiS collaboration), 1 AGeV $^{197}$Au$+^{12}$C, 1 AGeV $^{139}$La$+^{12}$C and 1 AGeV $^{84}$Kr$+^{12}$C (all performed by the EOS collaboration).\ \ The charge yields of all reactions as a function of excitation energy were fit with a version of Fisher's droplet model modified to account for the dual components of the fluid (i.e. protons and neutrons), Coulomb effects, finite size effects and angular momentum arising from the nuclear collisions.
\end{abstract}

\maketitle

\section{introduction}

In the long history of the study of the liquid to vapor phase transition of nuclear matter\cite{finn-82,siemens-83,hirsch-84,gilkes-94,pochodzalla-95,campi-97,moretto-97,bonasera-00,dagostino-00,elliott-00.2,elliott-02,berkenbusch-02,srivastava-02,dagostino-99} various studies have sought to determine one or more critical exponents \cite{finn-82,gilkes-94,elliott-00.2,elliott-02,berkenbusch-02,dagostino-99}, other studies have examined caloric curves \cite{pochodzalla-95}, and others have reported the observation of negative heat capacities \cite{dagostino-00}.\ \ All of these efforts suffer from the lack of knowledge of the systemÕs location in pressure-density-temperature ($p$, $\rho$, $T$) space.\ \ Specifically, interpretations of caloric curves and negative heat capacities depend on assumptions of either constant pressure or constant density \cite{moretto-96,elliott-00.3}.\ \ In the case of determining critical exponents, it was assumed that the systems were at coexistence and that the surface energy was the single, dominant factor.\ \ The analysis presented below makes no assumptions about the location of the system in ($p$, $\rho$, $T$) space and accounts specifically for other energetic considerations (e.g. the Coulomb force and angular momentum).

Our approach begins with the time honored idea of considering nuclei as drops of a hypothetical nuclear fluid.\ \ The liquid drop model \cite{weizsacker-35} takes up this idea quantitatively.\ \ The approximately constant binding energy per particle in heavier nuclei suggests that this fluid is bound together by a saturating short range force similar to that acting between the molecules of simple fluids (i.e. Van der Waals like).

Present day formulations of the liquid drop model \cite{myers-00,royer-06} express the binding energy in terms of a volume term proportional to the number of nucleons $A$ and corrected for finiteness by means of an expansion in terms of $A^{-1/3}$ of which only the first (surface energy) order term is kept.\ \ Additional corrections are added to account for neutron/proton asymmetry, Coulomb interactions and pairing effects.

Global fits to nuclear masses lead, on the one hand to a reproduction of binding energies to within 1\% ($\lesssim$10 MeV), and on the other to the characterization of the hypothetical fluid mentioned above, where finiteness, neutron/proton asymmetry and Coulomb have been removed.\ \ This is the ``bulk nuclear matter'' which has been studied theoretically over the history of nuclear physics \cite{bulk-paper}.

Van der Waals fluids admit various phases, among which are the vapor and the liquid.\ \ Thermodynamically, the equilibrium coexistence of these phases and the associated liquid-vapor phase transition are well understood.\ \ The Van der Waals aspects of the nuclear binding energy lead naturally to the question: does nuclear matter sustain a vapor phase as well as the condensed liquid phase?\ \ Is there in the phase diagram a (first order) coexistence line terminating at a critical point \cite{finn-82}?\ \ If so, how can one obtain such information experimentally?

We answer those questions (in the affirmative) and determine the coexistence curve and critical point of bulk nuclear matter from data obtained in three experiments and six different reactions.

In this paper, we describe first a physical picture of the nuclear reactions in question.\ \ Then we provide a brief description of the experiments.\ \ Greater detail about the experiments can be found in the references provided.\ \ Next, we give a detailed description of the theory and analysis used on the experimental data.\ \ Finally, we use the results of that analysis to determine the coexistence curve and critical point.

\section{The physical picture}

Thermal nuclear sources (compound nuclei and higher energy nuclear aggregates) emit particles such as neutrons, protons and heavier charged fragments into vacuum in a process that is very similar to evaporation \cite{wiesskopf-37}.\ \ This type emission from thermal, equilibrated systems is in contrast to the direct, or prompt, particle emission from excited nuclear systems out of equilibrium.

For fluid systems like water, evaporation rates allow one to recover the properties of the saturated vapor in equilibrium \cite{moretto-05}.\ \ However, in nuclear systems, finiteness and the presence of Coulomb effects prevent such a simple approach to the characterization of the phase diagram.

It has been shown previously how it is possible to ``reduce'' the nuclear evaporation rate to that of an infinite, uncharged symmetric fluid: finiteness is accounted for in terms of the ``complement'' approach \cite{moretto-05.1}, Coulomb effects can similarly be factored out \cite{moretto-03}, and the corrected rates can then be related to the properties of the hypothetical nuclear matter vapor \cite{moretto-05}.

We now demonstrate explicitly how properties of the bulk nuclear matter such as its phase diagram can be ascertained from experimental measurements of fragment distributions starting with a physical picture of fragment production from excited nuclei.\ \ This illustrates how one can talk about coexistence without the vapor being present and how an equilibrium description, such as Fisher's theory \cite{fisher-67.1,fisher-67.2} (described in section \ref{fisher-theory}), is relevant to the free vacuum decay of an evaporating, nuclear system \cite{moretto-05.1}.

Thermodynamicians would determine a phase diagram via direct measurements of the pressure, density and temperature of their fluid.\ \ However, such direct measurements of temperature, density and pressure for a nuclear fluid are problematic.\ \ On the other hand, the measurement of clusters in nuclear reactions has been easily achieved and has a long tradition.\ \ Thus, we believe that in nuclear physics, this is the royal avenue toward the extraction of the phase diagram.

\subsection{The virtual vapor}
\label{subsec:virtual}

Let us now consider a liquid in equilibrium with its saturated vapor.\ \ At equilibrium, any particle evaporated by the liquid is restored on the average by the vapor bombarding it.\ \ In other words, the outward evaporation flux from the liquid to the vapor is matched by the inward condensation flux.\ \ This is true for any kind of evaporated particle.\ \ The vapor acts as a mirror that reflects the evaporated particles back into the liquid.

One could probe the saturated vapor by putting a detector in contact with it.\ \ However, since the outward and inward fluxes are identically the same at equilibrium putting, the detector in contact with the liquid also probes the vapor.\ \ Therefore, we do not need the vapor to be physically present in order to characterize it completely.\ \ We can study the evaporation of the liquid and dispense with the surrounding saturated vapor.\ \ In these situations one thinks of a \emph{virtual vapor}, realizing that first order phase transitions depend exclusively upon the intrinsic properties of the two phases, and not on their interaction.\ \ Of course, if the vapor is not there to restore the emitting system with its return flux, evaporation will proceed and the system will cool.

An excited nucleus is a small drop of equilibrated nuclear matter that emits neutrons, protons and higher charged fragments into vacuum according to statistical decay rate theory.\ \ In this picture there is no surrounding vapor, no confining box and no need for either.\ \ As described in the preceding paragraph, by studying the outward flux of the first fragments emitted from a thermal source at equilibrium, we can study the nature of the vapor even when it is absent (the virtual vapor).

Quantitatively, the concentration $n_{A_{\text f}}(T)$ of any species $A_{\text f}$ constituents at temperature $T$ in the vapor is related to the corresponding decay rate $R_{A_{\text f}}(T)$ (or to the decay width $\Gamma _{A_{\text f}}(T)$) from the nucleus by matching the evaporation and condensation fluxes
\begin{equation}
	R_{A_{\text f}}(T) =\frac{\Gamma_{A_{\text f}}(T)}{\hbar} \approx n_{A_{\text f}}(T) \left< v_{A_{\text f}}(T) 4 \sigma _{\rm inv}(v_{A_{\text f}})\right>,
\label{eq:rate}
\end{equation}
where $v_{A_{\text f}}(T)$ is the thermal velocity of the species $A_{\text f}$ (of order $\sqrt{T/A_{\text f}}$) crossing the nuclear interface represented by the cross section $\sigma _{\rm inv}$ (of order $A_{\text s}^{2/3}$ where $A_{\text s}$ is the mass number of evaporating nucleus), $T$ is the temperature of the equilibrated, excited nucleus when the first fragment is emitted, $\hbar$ is Planck's constant and Coulomb effects have been, for the time being, neglected (they are dealt with below).

Equation~(\ref{eq:rate}) shows the fundamental and simple connection between the (compound nucleus) decay rate and the fragment concentration in the vapor.\ \ Thus, the vapor phase in equilibrium can be completely characterized in terms of the decay rate.

The physical picture described above is valid instantaneously.\ \ The result of successive evaporation in vacuum leads to abundances of various species of emitted fragments that arise from a continuum of systems at different temperatures \cite{ma-05}.\ \ This leads to complications in various thermometers: kinetic energy, isotope ratios, etc.

Our way of avoiding this complication is to consider only fragments that are emitted very rarely so that, if they are not emitted first, they are effectively not emitted at all.\ \ In other words, we consider only fragments that by virtue of their large surface energy (and high charge), have a high emission barrier.\ \ The rapidly increasing Coulomb barrier with fragment charge $Z$ strongly enhances this effect.\ \ Thus a lower cut-off of about $Z \approx 6$ is used in the analysis that follows.

\section{Experiments}

The above physical picture is used here to analyze the data from two kinds of experiments: compound nuclear decay and multifragmentation.\ \ Both types of experiments measure the total yield or number of fragments emitted from a thermal nuclear source, $Y_{Z_{\text f}} \left( E^*_{\text s} \right)$, as a function of the excitation energy of the source, $E^*_{\text s}$, and charge of the fragment, $Z_{\text f}$.\ \ For both types of  experiments it is assumed that the collisions produce an excited, equilibrated thermal source of radius $r_{\text s}$ consisting of $A_{\text s}$ nucleons ($Z_{\text s}$ protons and $N_{\text s}$ neutrons) at excitation energy $E^{*}_{\text s}$ and with angular momentum $\vec{I}$.\ \ This is the initial state of the system: an excited, thermal nucleus which emits neutrons, protons and heavier charged fragments.

\subsection{Compound nucleus}

The first kind of experiment gives rise to a \emph {compound nucleus} \cite{cn-rev}.\ \ A compound nucleus is formed when one nucleus impacts another nucleus and the two combine to form a single, compound system.\ \ The nucleon number and charge of the compound nucleus is just the sum of the nucleon number and charge of the two colliding nucleii.\ \ Its excitation energy can be determined from the energy of the bombarding nuclei and the masses of the target and projectile.\ \ The excited compound nucleus is a thermal source that emits protons, neutrons and other heavier charged fragments.

The compound nucleus experiments analyzed here were performed at the 88" Cyclotron of the Lawrence Berkeley National Laboratory \cite{fan-00}.\ \ An Advanced Electron-Cyclotron-Resonance (AECR) ion source \cite{xie-91} was utilized to produce highly charged $^{58}$Ni and $^{64}$Ni ions which, after injection into the cyclotron and acceleration to the desired energy, impinged on a high purity \cite{mcmahan-86} carbon target (1.0 mg/cm$^2$).\ \ The fragments emitted in the reactions were detected in two position-sensitive $\Delta E - E$ detector assemblies placed on either side of the beam.\ \ The methods of the energy and position calibrations of the $\Delta E$ and $E$ detectors have been described previously \cite{mcmahan-86,jing-99}.

\begin{figure}[htbp]
\includegraphics[width=8.0cm]{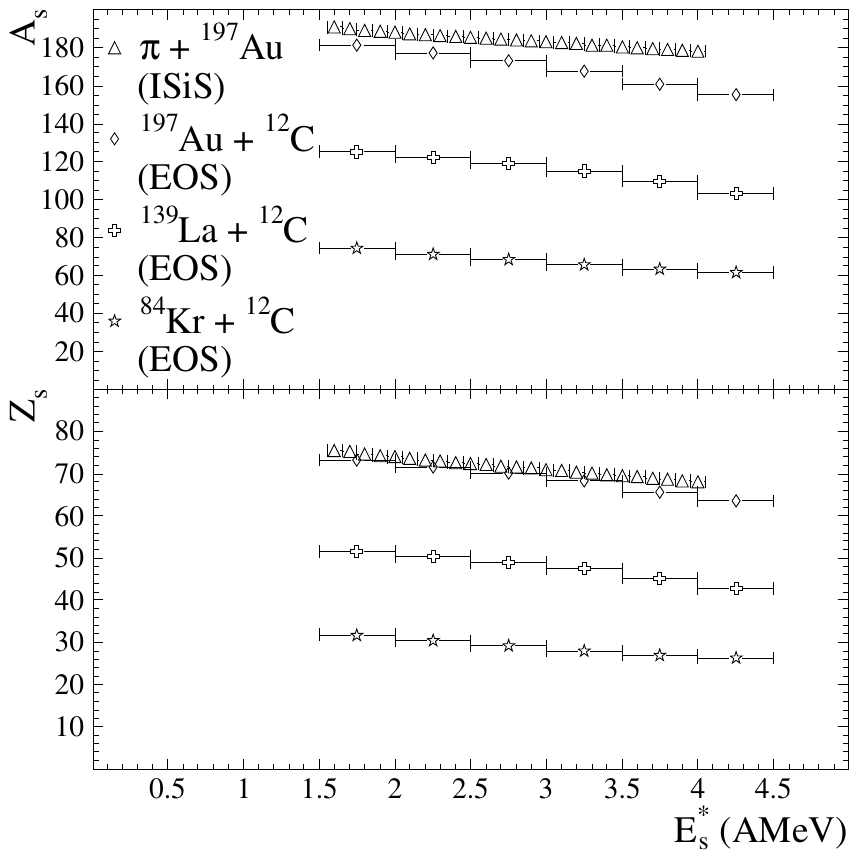}
\caption{Top: the nucleon number as a function of excitation energy for the thermal sources created by the multifragmentation experiments.\ \ Bottom: the charge of the thermal source as a function of excitation energy.}
\label{remnant-az}
\end{figure}

\subsection{Nuclear multifragmentation}

The second kind of experiment analyzed in this work gives rise to a phenomenon called \emph {multifragmentation} \cite{aichlen-91,bondorf-95,gross-97,moretto-97,richert-01,chomaz-04,das-05,viola-06}.\ \ In a multifragmentation experiment, one nucleus is accelerated to a high velocity and impacts another nucleus and in the experiments considered here, one of the colliding nuclei is larger than the other.\ \ Typically, the collision between nuclei in multifragmentation experiments is more violent than the that in compound nucleus experiments.\ \ The two nuclei either partially fuse or a ``fireball'' is generated from the occluded parts of the target and projectile.\ \ The larger of the two nuclei promptly loses nucleons during the collision leading to an excited, thermal remnant with a smaller nucleon number and charge than the initial nuclei.\ \ In the experiments considered here, the greater the excitation energy of the remnant, the the smaller the nucleon number and charge of the remnant.\ \ Figure~\ref{remnant-az} shows the nucleon number, $A_{\text s}$, and charge, $Z_{\text s}$, as a function of excitation energy, $E^{*}_{\text s}$, for the remnants created in the multifragmentation experiments considered here.\ \ The excited remnant is a thermal source that emits protons, neutrons and other heavier charged fragments.\ \ In multifragmentation experiments, the excitation energy is estimated based on measurements of the kinetic energy of the fragments emitted from the remnant and other considerations \cite{hauger-96, hauger-00}.

\subsubsection{EOS}

Some of the multifragmentation data analyzed here are from the reactions 1 AGeV $^{197}$Au$+^{12}$C, 1 AGeV $^{139}$La$+^{12}$C and 1 AGeV $^{84}$Kr$+^{12}$C and were collected by the EOS Collaboration at the Lawrence Berkeley National Laboratory Bevalac.\ \ This experiment studied the projectile fragmentation and detected nearly all of the charged reaction products on an event-by-event basis \cite{hauger-96,hauger-00,gilkes-94,elliott-03}.\ \ Charged particles with charges of from $1$ to $6$ were identified using a time projection chamber \cite{rai-90} while a multiple sampling ionization chamber detected charged particles with charges from $7$ to $79$ \cite{christie-87}.

\subsubsection{ISiS}

The other multifragmentation data analyzed here are from the reaction 1 GeV$/$c $\pi+^{197}$Au and were collected by the ISiS Collaboration at the Alternating Gradient Synchrotron (AGS) at Brookhaven National Laboratory \cite{viola-06,kwiatkowski-95, lefort-99, beaulieu-00,beaulieu-01}.\ \ The AGS provided beams of 1 GeV$/$c $\pi$ incident on a gold target.\ \ Particles with charges from $1$ to $16$ were measured by the Indiana Silicon Sphere (ISiS) $4\pi$ detector array  \cite{kwiatkowski-95} providing a high statistics, exclusive data set with finer bins in excitation energy than the EOS experiment \cite{lefort-99}.

\section{Analysis}

We now provide a derivation of a formula for the average fragment yields that should be observed in the experiments based on the initial state of the system (the excited, compound nucleus or remnant) and the final state of the system (the fragment and its complement).\ \ We start from Fisher's droplet model \cite{fisher-67.1,fisher-67.2} and modify it to account for effects that arise from finite size \cite{moretto-05.1}, the Coulomb force \cite{moretto-03}, isospin, angular momentum and the secondary decay of excited fragments. 

\subsection{Fisher's droplet model and the complement}
\label{fisher-theory}

Fisher's droplet model \cite{fisher-67.1,fisher-67.2} is a physical cluster theory that has successfully: described the cluster distributions in percolating systems \cite{elliott-03} and lattice gas (Ising) systems \cite{mader-03}; reproduced the compressibility factor at the critical point \cite{kiang-70}; predicted (within a few percent) the compressibility factor of real fluids from the triple point to the critical temperature \cite{rathjen-72,saltz-94};  and has been used to describe the nucleation rate of real fluids \cite{stauffer-72,dillmann-91}.

Physical cluster theories of non-ideal fluids assume that the monomer-monomer interaction is exhausted in the formation of clusters, and that the resulting clusters behave ideally (i.e. they do no interact with each other) \cite{mayer-40,frenkel-46}.\ \ Further, clusters of a given number of constituents $A_{\rm f}$ can be characterized by a chemical potential (per constituent) $\mu$ and a partition function $q_{A_{\rm f}} (T,V)$ that depends on the temperature $T$ and volume $V$ of the fluid and is given by
\begin{eqnarray}
	q_{A_f}(T,V) & = & V \left( \frac{ 2 \pi m_{A_f} T}{h^2} \right)^{\frac{3}{2}}  \exp\left( -\frac{\Delta G}{T} \right) 
\label{cluster-part-fcn-1}
\end{eqnarray}
where $V$ is the volume, $m_{A_{\rm f}}$ is the mass of a fragment of $A_{\rm f}$ constituents.\ \ Here $\Delta G$ is the free energy cost for the formation of that cluster \cite{stauffer-77}
\begin{equation}
\Delta G = \Delta E - T \Delta S + p \Delta V
\label{deltag}
\end{equation}
where $\Delta E$ and $\Delta S$ are the energy and entropy cost of the formation of the cluster respectively and $p$ is the pressure and $\Delta V$ is the change in volume due to the formation of the cluster.

Because of the ideality of the fluid of clusters, the pressure and density are readily determined. The pressure $p$ is
\begin{equation}
	p = \frac{T}{V} \sum_{A_{\rm f}=1}^{\infty} q_{A_{\rm f}} (T,V) z^{A_{\rm f}}
\label{pressure-01}
\end{equation}
and the density $\rho$ is
 \begin{equation}
	\rho = \frac{1}{V}\sum_{A_{\rm f}=1}^{\infty} A q_{A_{\rm f}} (T,V) z^{A_{\rm f}}
\label{density-01}
\end{equation}
where $z$ is the fugacity $z=e^{\mu/T}$. The concentration of size $A_{\text f}$ clusters is then 
\begin{eqnarray}
	n_{A_{\rm f}} (T,z) &=& \frac{q_{A_{\rm f}}(T,V) z^{A_{\rm f}}}{V} \nonumber \\
	& = & z^{A_{\rm f}} \left( \frac{2 \pi m_{A_{\rm f}} T}{h^2} \right)^{\frac{3}{2}} \exp \left( - \frac{\Delta G}{T} \right) \nonumber \\
	& = & q_0 \exp \left( \frac{A_{\rm f} \Delta \mu}{T} \right) \exp \left( - \frac{\Delta G}{T} \right)
\label{cluster-conc-01}
\end{eqnarray}
where $\Delta \mu$  is a measure of the distance from coexistence in terms of the chemical potential which, following Fisher \cite{fisher-67.1}, absorbs the thermal wavelength and $q_0$ is a normalization constant.\ \ At coexistence, $\Delta \mu = 0$, the cluster concentration is given by
\begin{equation}
n_{A_{\rm f}}(T) = q_0 \exp \left(-\frac{\Delta G}{T} \right) .
\label{fisher-00}
\end{equation}

There have been many derivations of $\Delta G$, but here we follow a general derivation using the complement method \cite{moretto-05.1} and concentrate on the change in free energy between the initial state and the final state.\ \ Because in the complement derivation bulk terms do not survive, they are omitted in the derivation below.

The initial state consists of an equilibrated liquid drop consisting of $A_{\rm s}$ particles.\ \ In the final state the drop has just emitted a cluster or droplet or fragment with $A_{\rm f}$ particles.\ \ Also in the final state along with the fragment is the complement.\ \ The complement is what is left of the drop after the fragment has been emitted and thus consists of $A_{\rm c} = A_{\rm s} - A_{\rm f}$ particles.

In determining the free energy of the initial and final state, we follow Fisher's contribution to physical cluster theory which was to endow clusters with a surface energy and to provide an estimate for the entropic part of the free energy associated with the formation of a cluster \cite{fisher-67.1,fisher-67.2}.

Since the vapor of clusters is ideal, its internal energy is given by
\begin{equation}
E =  \sum_{A=1}^{A=A_{\rm s}} n_A E_A
\label{Eint}
\end{equation}
where $E_A$ is the binding energy of a cluster and is determined via a liquid-drop expansion
\begin{equation}
E_A = a_v A + a_s A^{\sigma} 
\label{EA}
\end{equation}
where $a_v$ is the bulk of volume energy coefficient, $a_s$ is the surface energy coefficient and $\sigma$ is an exponent describing the relationship between the surface and volume of the fragment.\ \ One intuitively expects that $\sigma \approx 2/3$ for three dimensional systems.

The only contributions to the change in the energy between the final and initial states are those that are associated with the change in the net surface area.\ \ Thus the energy change is given by
\begin{equation}
\Delta E = a_s \left[ \left( A_{\rm s} - A_{\rm f} \right)^{\sigma} + A_{\rm f}^{\sigma} - A_{\rm s}^{\sigma} \right] .
\label{DE}
\end{equation}

Similarly, the entropy of the fluid is given by
\begin{equation}
S = \sum_{A=1}^{A=A_{\rm s}} n_A S_A .
\label{Sint}
\end{equation}
Fisher conjectured that the entropy of a cluster could also be estimated with a liquid drop type expansion
\begin{equation}
S_A = b_v A + b_s A^{\sigma} - \tau \ln A
\label{SA}
\end{equation}
where $b_v$ is the bulk of volume entropy coefficient and $b_s$ is the surface entropy coefficient.\ \ See reference  \cite{fisher-59} for more on the origins of the logarithmic term.

As with the energy, the only contributions to the change in the entropy between the final and initial states are those associated with the change in the net surface area.\ \ Thus the entropic contribution is given by 
\begin{eqnarray}
\Delta S &=& b_s \left[ \left( A_{\rm s} - A_{\rm f} \right)^{\sigma} + A_{\rm f}^{\sigma} - A_{\rm s}^{\sigma} \right] \nonumber \\
&-& \tau \ln \left[ \frac{\left( A_{\rm s} - A_{\rm f} \right)A_{\rm f}}{A_{\rm s}}\right] .
\label{DS}
\end{eqnarray}

The contribution to the free energy due to the change in volume, the $p\Delta V$ term is negligible compared to the energetic and entropic contributions and will be ignored here.

Equations~(\ref{Eint}) and (\ref{Sint}) combine to show that the change in free energy is
\begin{eqnarray}
\Delta G & = & \left( a_s - T b_s \right)  \left[ \left( A_{\rm s} - A_{\rm f} \right)^{\sigma} + A_{\rm f}^{\sigma} - A_{\rm s}^{\sigma} \right] \nonumber \\
& + & T \tau \ln \left[ \frac{\left( A_{\rm s} - A_{\rm f} \right)A_{\rm f}}{A_{\rm s}}\right] .
\label{Gint}
\end{eqnarray}

At the critical temperature, $T_c$, the surface's contribution to $\Delta G$ vanishes leaving only the logarithmic term, thus eq.~(\ref{Gint}) shows that $T_c$ is  defined as
\begin{equation}
	T_c = \frac{a_s}{b_s} .
\label{crit-temp}
\end{equation}
Assuming that the coefficients $a_s$ and $b_s$ are independent of the temperature the quantity $\left( a_s - T b_s \right)$ can be rewritten as $a_s \varepsilon$ with
\begin{equation}
	\varepsilon = \frac{T_c - T}{T_c} .
\label{eps}
\end{equation}

Now the fragment concentration becomes 
\begin{eqnarray}
n_{A_{\rm f}}(T) & = & q_0 \left[ \frac{\left( A_{\rm s} - A_{\rm f} \right)A_{\rm f}}{A_{\rm s}}\right]^{\tau} \nonumber \\
& \times & \exp \left\{ -\frac{ a_s \varepsilon \left[ \left( A_{\rm s} - A_{\rm f} \right)^{\sigma} + A_{\rm f}^{\sigma} - A_{\rm s}^{\sigma} \right]}{T} \right\}
\label{fisher-00a}
\end{eqnarray}

In the limit of a bulk liquid where $A_{\rm s} \rightarrow \infty$, then $A_{\rm s}- A_{\rm f} \approx A_{\rm s}$ and we obtain
\begin{equation}
n_{A_{\text f}} \left( T \right) = q_0 A_{\text f}^{-\tau} \exp \left( -\frac{a_s \varepsilon A_{\text f}^{\sigma}}{T} \right) 
\label{fisher-01}
\end{equation}
which is the expression Fisher derived for the cluster concentration \cite{fisher-67.1,fisher-67.2,stauffer-77}

At the critical point, $\varepsilon=0$ (the surface free energy vanishes) and the fragment distribution is given by a power law.\ \ The power law has been explicitly verified in percolation and Ising systems \cite{stoll-72,stauffer-79,cambier-86,bauer-88,kertesz-89,wang-89,wang-90,elliott-94,latora-94.1,latora-94.2,pratt-95,belkacem-95,finocchiaro-96,dasgupta-96.2,campi-97,strachan-97,elliott-97.2,kondratyev-97,carmona-98,strachan-99,gulminelli-99,dorso-99,elliott-00.1,elliott-00.2,elliott-03,mader-03,elliott-05} and implicitly verified in a wide variety of physical fluids \cite{kiang-70,rathjen-72}.

For the present work, the liquid is not infinite and at most $A_{\rm s} \approx 175$, so equation~(\ref{fisher-01}) cannot be used and all the terms in equation~(\ref{fisher-00a}) must be used, in addition to other terms that arise due to the nuclear nature of the fluid, e.g. a Coulomb term, an isospin term, etc.\ \ Those other terms are derived below by examining the initial and final states of the evaporating, equilibrated, excited nuclear source.

\subsection{Characterization of the initial state: Properties of the thermal source}

The nucleon number of the excited, equilibrated, evaporating nuclear source is $A_{\text s}$ with $Z_{\text s}$ protons and $N_{\text s}$ neutrons.\ \ For the compound nucleus experiments, the source is defined as the sum of the target and projectile.\ \ For the multifragmentation experiments the source was measured in the experiments and found to be a nucleus smaller in nucleon number than the larger of the projectile or target as shown in Figure~\ref{remnant-az}.

The temperature of the thermal source $T$ is estimated via the Fermi gas.\ \ The excitation energy of the source in terms of MeV per nucleon is related to the temperature of the source by
\begin{equation}
E^{*}_{\text s} = a T^2
\label{temp}
\end{equation}
where the level density parameter, $a$, is modified to account for the empirically observed change with excitation energy \cite{hagel-88} and is given by \cite{raduta-97}
\begin{equation}
\frac{1}{a} = 8 \left[ 1 + \left( \frac{A_{\text s}E^{*}_{\text s}}{E^{\text {bind}}_{\text s}} \right) \right] .
\label{lev-den}
\end{equation}

In Eq.~(\ref{lev-den}) $E^{\text{bind}}_{\text s}$ is the binding energy of the thermal source.\ \ Because the fragment yield distributions analyzed below are measured as a function of fragment charge ($Z_{\text f}$) and because the fragment mass ($A_{\text f}$) is only estimated, pairing and shell effects are neglected and the binding energy of a nucleus of nucleon number $A$ ($Z$ protons and $N$ neutrons) is found (in MeV) from the liquid drop expansion \cite{weizsacker-35,myers-00,royer-06}
\begin{eqnarray}
\label{Ebind}
E^{\text {bind}}_{A,Z} & = &  -a_v \left( 1 -k_v y^2 \right) A \nonumber \\
& + & a_s\left( 1 - k_s y^2 \right) A^{\sigma}  \\
& + & \kappa \frac{3}{5} \frac{Z \left( Z - 1 \right)}{r_0 A^{1/3}} \nonumber
\end{eqnarray}
where $a_v = 15.7335$, $k_v = 1.6949$, $a_s = 17.8048$,  $k_s = 1.0884$,  $\kappa=1.43997$ MeV fm, $r_0 = 1.2181~\text{fm}$ and the asymmetry (or isospin) parameter is
\begin{equation}
y = \frac{A-2Z}{A} .
\label{asym-para}
\end{equation}
The values of the parameters are taken from reference \cite{royer-06}.

We note that Eq.~(\ref{Ebind}) already gives us $a_s$, one of the parameters needed to determine the critical temperature as shown in Eq.~(\ref{crit-temp}).

In the case of the compound nucleus experiments, the angular momentum $\vec{I}$ of the thermal source was estimated \cite{fan-00}.\ \ However, we multiplied that estimate by a constant, $I_0$, that was left as a fitting parameter.\ \ For the multifragmentation experiments, we parametrized $\vec{I}$ as
\begin{equation}
\left| \vec{I} \right| = \left| I_0 + I_1 E^{*}_{\text s} \right|
\label{poly-l}
\end{equation}
where the coefficients $I_0$ and $I_1$ are also left as fit parameters

The energy of the thermal source due to its angular momentum is (classically)
\begin{equation}
E^{\vec{I}}_{\text s} = \frac{\left|\vec{I}\right|^2}{\frac{4}{5}m_{\text s} r^{2}_{\text s}} .
\label{Erangmo-r}
\end{equation}
All radii in this work are taken to be
\begin{equation}
r = r_0 A^{1/3} .
\label{rad0}
\end{equation}

Following Fisher, the entropy of a nucleus is estimated based on Eq.~(\ref{SA}) with $\sigma$ and $\tau$ set to their three dimensional Ising class value: $\sigma = 0.63946 \pm 0.0008$ and $\tau = 2.209\pm0.006$ \cite{elliott-03}.

The free energy of the thermal source is the free energy of the initial state
\begin{eqnarray}
 G_{\text {initial}} & = & E^{\text {bind}}_{\text s}+ E^{\vec{I}}_{\text s} - T S_{\text s} .
\label{Frem}
\end{eqnarray}
The $p V$ contribution to the free energy is neglected here.

\subsection{Final state: the fragment and complement}

The final state considered here is immediately after the emission (or evaporation) of a neutron, proton or heavier charged fragment.\ \ The fragment has mass $m_{\text f}$, radius $r_{\text f}$ and $A_{\text f}$ nucleons ($Z_{\text f}$ protons and $N_{\text f}$ neutrons).

After fragment emission the thermal source is reduced in nucleon number and labeled as the ``complement'' and has mass $m_{\text c}$, radius $r_{\text c}$ and $A_{\text c}$ nucleons ($Z_{\text c}$ protons and $N_{\text c}$ neutrons).\ \ It is assumed that the the fragment and complement are both spherical, at normal nuclear density and that the surface of the fragment and the surface of the complement are in contact.

Conservation of angular momentum dictates that the fragment and complement system have the same angular momentum as the thermal source.\ \ The energy associated with this angular momentum is then (classically)
\begin{equation}
E^{\vec{I}}_{\text {f+c}} = \frac{\left|\vec{I}\right|^2}{\frac{4}{5} \left( m_{\text c} r^{2}_{\text c} +  m_{\text f} r^{2}_{\text f} \right) + 2 \frac{m_{\text c} m_{\text f}}{m_{\text c} + m_{\text f}} \left( r_{\text c} + r_{\text f} \right)^2} . 
\label{Erangmo-cf}
\end{equation}

The Coulomb energy between the fragment and complement is
\begin{equation}
E^{\text {Coulomb}}_{\text {f+c}} = \kappa  \frac{Z_{\text f} Z_{\text c}}{r_{\text f} + r_{\text c}} .
\label{ECoul}
\end{equation}

The free energy of the final state is
\begin{eqnarray}
G_{\text {final}} & = & E^{\text {bind}}_{\text f} + E^{\text {bind}}_{\text c} + E^{\vec{I}}_{\text {f+c}} + E^{\text {Coulomb}}_{\text {f+c}} \nonumber \\
&- &  T \left( S_{\text f} + S_{\text c} \right) .
\label{Ffinal}
\end{eqnarray}
The determination of quantities associated with the fragment and complement are discussed below.

\subsubsection{Properties of the fragment}

The charge of the fragment $Z_{\text f}$ is measured by the detectors in the experiments.\ \ The nucleon number of the observed fragment $A_{\text f}^{\text {EPAX}}$ is estimated via the EPAX parameterization \cite{summerer-90}
\begin{equation}
A_{\text f}^{\text {EPAX}} = -0.10167 + 1.9638Z_{\text f} + 0.0057221Z_{\text f}^2 .
\label{epax}
\end{equation}

The excitation energy of the fragment is estimated with the Fermi gas relation
\begin{equation}
E^{*}_{\text f} = a {A_{\text f}} T^2 = a \left({A_{\text f}^{\text {EPAX}}+N_{\text {evap}}}\right) T^2
\label{fragE*}
\end{equation}
where $A_{\text f}$ is the nucleon number of the fragment prior to any secondary decay or evaporation (under the approximation that only neutrons are evaporated from the fragments).

The number of neutrons $N_{\text {evap}}$ that can be evaporated from a nucleus $\left( A_{\text f}^{\text{EPAX}}, Z_{\text f} \right)$ is approximately:
\begin{equation}
N_{\text {evap}} \approx \frac{E^{*}_{\text f}}{B_n + 2T}
\label{nevap}
\end{equation}
where $B_n$ is the neutron binding energy of the nucleus in question.\ \ This is estimated as
\begin{eqnarray}
B_n & \approx & m_{A_{\text f}^{\text {EPAX}}-1, Z_{\text f}} + m_{\text{n}} - m_{A_{\text f}^{\text {EPAX}}, Z_{\text f}} \nonumber \\ 
& \approx & E^{\text {bind}}_{A_{\text f}^{\text {EPAX}}-1, Z_{\text f}} -E^{\text {bind}}_{A_{\text f}^{\text {EPAX}}, Z_{\text f}} .
\label{Ebn}
\end{eqnarray}

Combining Equations~(\ref{fragE*}) and (\ref{nevap}) gives
\begin{equation}
N_{\text {evap}} \approx d_2 \left( \frac{ a{T^2} A_{\text f}^{\text {EPAX}}}{B_n + 2T - a{T^2}} \right) .
\label{origA1}
\end{equation}
where $d_2$ is a fit parameter to account for the crude nature of this approximation.

The fragment's initial nucleon number $A_{\text f}$ is
\begin{equation}
A_{\text f} = A_{\text f}^{\text {EPAX}} + N_{\text {evap}}.
\label{origA2}
\end{equation}
This estimate assumes that only neutrons are emitted during the secondary decay.

The fragment's binding energy ($E^{\text {bind}}_{\text f}$), radius ($r_{\text f}$) and entropy ($S_{\text f}$) are determined using equations~(\ref{Ebind}) and (\ref{SA}), respectively, with $A_{\text f}$ and $Z_{\text f}$.

\subsubsection{Properties of the complement}

Conservation of mass and charge give the mass $A_{\text c}$, charge $Z_{\text c}$ and neutron $N_{\text c}$ number of the complement.\ \ The binding energy ($E^{\text {bind}}_{\text c}$), radius ($r_{\text c}$) and entropy ($S_{\text c}$) of the complement are determined using equations~(\ref{Ebind}) and (\ref{SA}), respectively, with $A_{\text c}$ and $Z_{\text c}$.

\subsection{Average fragment yields}

All of the experiments discussed here measure the average yield of fragments with a given charge as a function of excitation energy of the thermal source $Y_{Z_{\text f}} \left( E^*_{\text s} \right)$ where:
\begin{equation}
Y_{Z_{\text f}} \left( E^*_{\text s} \right) = \frac{\text{number of } Z_{\text f} \text{ fragments in events with } E^*_{\text s}}{\text{total number of events with }E^*_{\text s} } . \nonumber
\label{yz-ave}
\end{equation}
and its associated error on the mean $\delta Y_{Z_{\text f}} \left( E^*_{\text s} \right)$.

It is assumed that there is a one to one relationship between the excitation energy of the thermal source and the temperature and a one to one relationship between the charge of a fragment and its nucleon number.\ \ Thus the fragment charge yields as a function of excitation energy are equivalent to the fragment mass yields as a function of temperature which will be written as $Y_{A_{\text f}, Z_{\text f}} \left( T \right)$.

For the first fragments emitted their yield is given by:
\begin{equation}
 Y_{A_{\text f}, Z_{\text f}} \left( T \right) = \Delta t R_{A_{\text f}, Z_{\text f}} \left( T \right) = Y
\label{yields-1}
\end{equation}
where $\Delta t$ is the time duration of the measurement of the decay of the thermal source and $R_{A_{\text f}, Z_{\text f}} \left( T \right)$ is the rate of fragment emission.

Equation~(\ref{eq:rate}) shows that the fragment emission rate from the thermal source is related to the concentration $n_{A_{\text f}, Z_{\text f}} \left( T \right)$ of any species $\left( A_{\text f}, Z_{\text f}, \right)$ in the ``virtual'' vapor which matches the evaporation (or emission) flux out of the thermal source with the condensation flux into the thermal source \cite{moretto-05}
\begin{equation}
\label{rate}
	R_{A_{\text f}, Z_{\text f}} \left( T \right)  \approx \left< v_{A_{\text f}} \sigma _{\rm inv}\right>  n_{A_{\text f}, Z_{\text f}} \left( T \right) .
\end{equation}

The mean thermal velocity of the fragment normal to the plane of emission is given as
\begin{equation}
\left< v_{A_{\text f}} \right> = \sqrt{\frac{T}{2 \pi m_{A{\text f}}}}
\label{vfrag}
\end{equation}
where an ideal vapor has been assumed.

The inverse cross section for fragment emission is
\begin{equation}
\sigma_{\text {inv}} = 4 \pi \left( r_{\text f} + r_{\text c} \right)^2 
\label{inv-cs}
\end{equation}
where only the geometric cross section is considered since the Coulomb effects are explicitly dealt with below \cite{moretto-03}.

Equation~(\ref{fisher-00}) shows that the concentration of the virtual vapor depends on the free energy cost of cluster formation $\Delta G$.\ \ For a fragment emitted from an excited nucleus $\Delta G$ is given by equations~(\ref{Frem}) and (\ref{Ffinal}) and is
\begin{eqnarray}
\label{DGAZ}
\Delta G & = & E^{\text {bind}}_{\text f} + E^{\text {bind}}_{\text c} + E^{\vec{I}}_{\text {f+c}} + E^{\text {Coulomb}}_{\text {f+c}}  - E^{\text {bind}}_{\text s} - E^{\vec{I}}_{\text {s}} \nonumber \\
& - & T \left\{ b_s \left ( A_{\text f}^{\sigma} + A_{\text c}^{\sigma} - A_{\text s}^{\sigma}\right) - \tau \ln \left( \frac{A_{\text f}A_{\text c}}{A_{\text s}} \right) \right\} .
\end{eqnarray}
This can be simplified and written as:
\begin{eqnarray}
\Delta G & = & G_{\text {final}} - G_{\text {initial}} \nonumber \\
& = & a_s A_{\text f}^{\sigma} - T \left( b_s A_{\text f}^{\sigma} - \tau \ln A_{\text f} \right) + \Delta \mu_{\text {nfs}} \nonumber \\
& = & \Delta G_{\infty} + \Delta \mu_{\text {nfs}}
\label{DGbulk}
\end{eqnarray}
where $ \Delta \mu_{\text {nfs}}$ absorbs all the terms in eq.~(\ref{DGAZ}) (and is divided by $A_{\text f}$) not explicitly written in eq.~(\ref{DGbulk}).\ \ $\Delta \mu_{\text {nfs}}$ is an ``effective chemical potential'' that arises due to the nuclear nature and finite size of the thermal source.\ \ The factors not absorbed in $ \Delta \mu_{\text {nfs}}$ describe the free energy cost of the formation of a fragment from bulk nuclear matter which is written as $\Delta G_{\infty}$.

\subsection{Fitting the experimental charge yields}

\begin{table}[htdp]
\caption{Fit details}
\begin{center}
\begin{tabular}{|ccccc|}
\hline
Reaction									& Points	& Number of		& $Z_{\text f}$		& $E^{*}_{\text s}$ (AMeV)	\\
										& fit		& Parameters	& Range				& Range					\\
\hline
$^{58}$Ni$+^{12}$C$\rightarrow^{70}$Se	&$54$	&	$3$			&$\left[ 6, 16 \right]$	&$\left[ 1.13, 2.02 \right]$	\\
$^{64}$Ni$+^{12}$C$\rightarrow^{76}$Se	&$40$	&	$3$			&$\left[ 7, 15 \right]$	&$\left[ 1.08, 1.82 \right]$	\\
1 AGeV $^{84}$Kr$+^{12}$C				&$26$	&	$4$			&$\left[ 6, 13 \right]$	&$\left[ 1.75, 4.75 \right]$	\\
1 AGeV $^{139}$La$+^{12}$C				&$53$	&	$4$			&$\left[ 6, 18 \right]$	&$\left[ 1.75, 4.75 \right]$	\\
1 AGeV $^{197}$Au$+^{12}$C				&$96$	&	$4$			&$\left[ 6, 25 \right]$	&$\left[ 1.75, 4.75 \right]$	\\
1 GeV$/$c $\pi+^{197}$Au					&$234$	&	$4$			&$\left[ 6, 15 \right]$	&$\left[ 1.50, 4.00 \right]$	\\
\hline
\end{tabular}
\end{center}
\label{summary}
\end{table}

\begin{figure}[htbp]
\includegraphics[width=8.0cm]{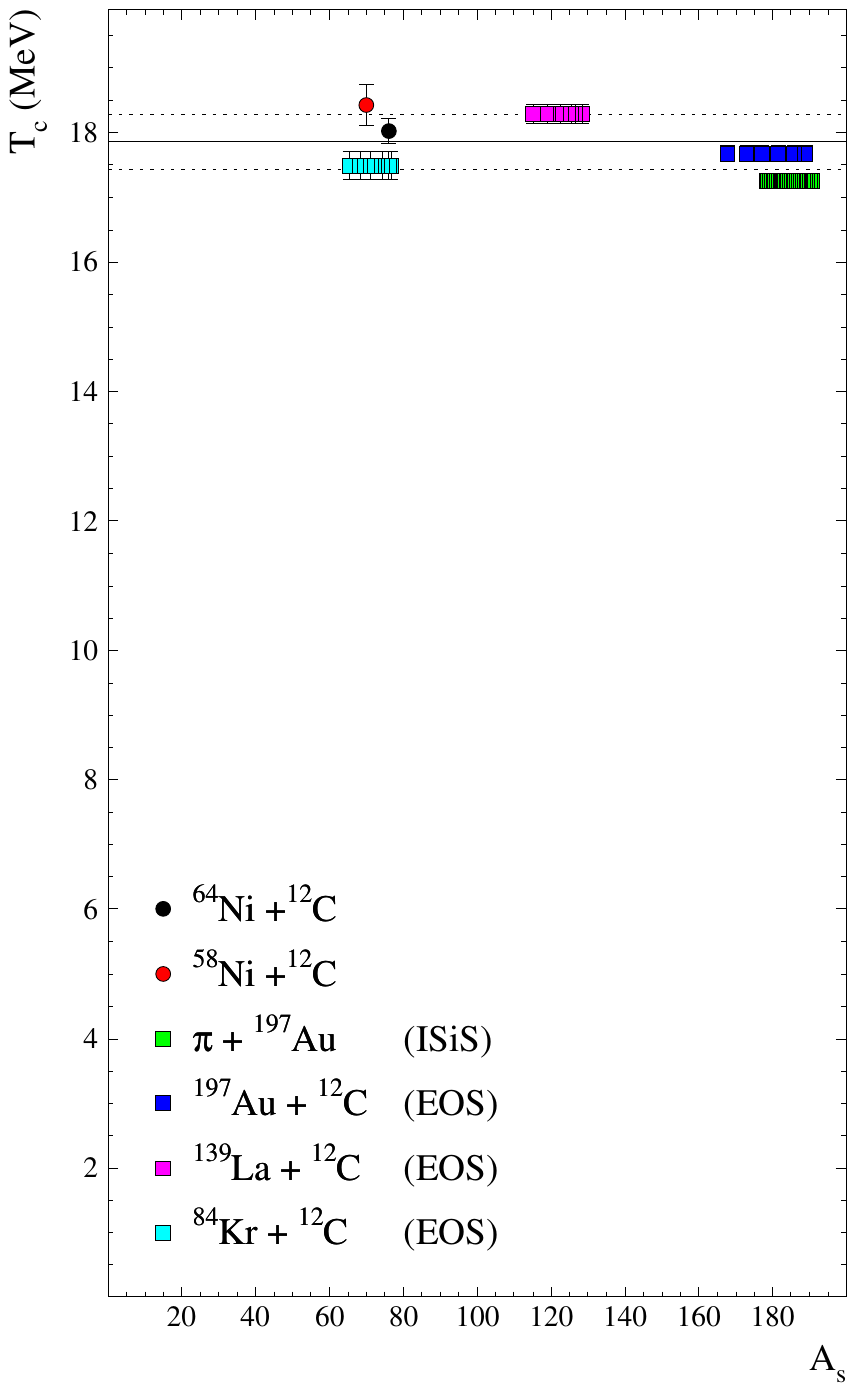}
\caption{The critical temperature as a function of the thermal source mass.\ \ The results of the compound nuclear reactions are shown with circles and the results of the multifragmentation reactions are shown with squares.\ \ Colors show the results for different experiments.\ \ The solid line shows the average of all the measurements and the dotted lines show the RMS variation: $17.9 \pm 0.4$ MeV.}
\label{tc}
\end{figure}

\begin{figure}[htbp]
\includegraphics[width=8.0cm]{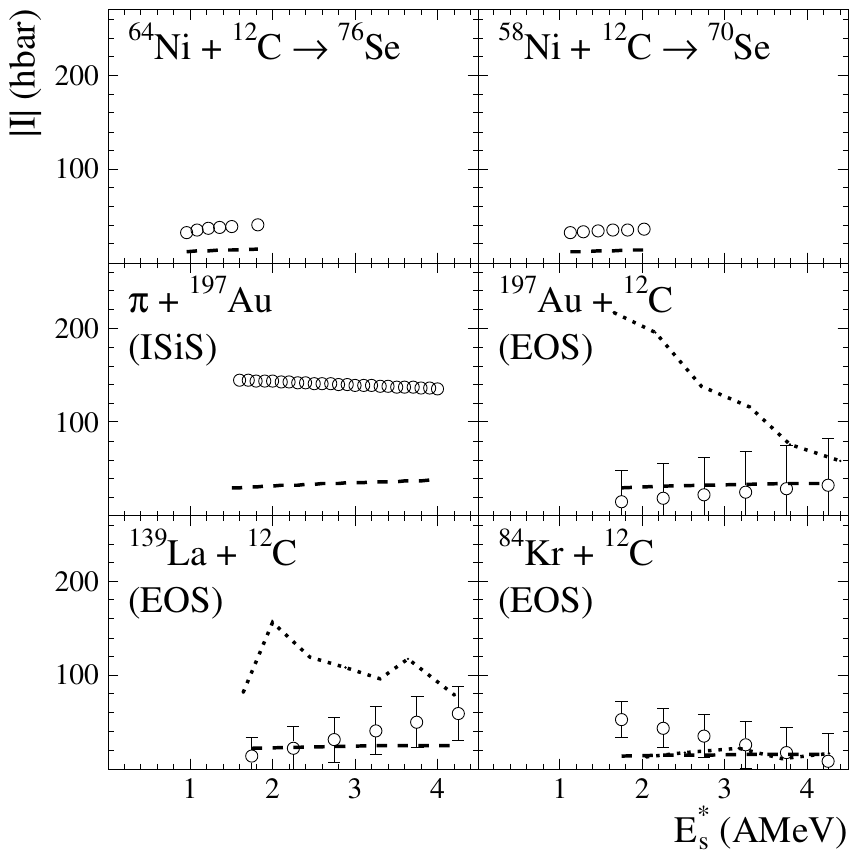}
\caption{The angular momentum values for all six reactions predicted by the fitting of the charge yields.\ \ Error bars are smaller than the points for the compound nucleus and ISiS experiments.\ \ The dashed line shows a thermal estimate of the angular momentum \cite{moretto-??}.\ \ Dotted lines show an estimate of the angular momentum based on the colliding nuclei in the multifragmentation experiments.\ \ See text for details.}
\label{angmo-plots}
\end{figure}

\begin{table*}[htdp]
\caption{Fitting results}
\begin{center}
\begin{tabular}{|ccccccc|}
\hline
Reaction								&	$\chi^{2}_{\nu}$	&$d_2$				& $I_0$			& $I_1$		& $b_s$			&	$T_c$  (MeV)		\\
\hline
$^{58}$Ni$+^{12}$C$\rightarrow^{70}$Se	&	$1.3$			&$0.1\pm0.1$			& $1.20\pm0.09$	&	-		& $0.97\pm0.02$	&	$18.4\pm0.3$		\\
$^{64}$Ni$+^{12}$C$\rightarrow^{76}$Se	&	$0.4$			&$0.5\pm0.2$			& $1.3\pm0.2$		&	-		& $0.99\pm0.01$	&	$18.0\pm0.2$		\\
1 AGeV $^{84}$Kr$+^{12}$C				&	$3.3$			&$0.0\pm5\times 10^{-5}$	& $-83\pm13$		& $18\pm4$	& $1.02\pm0.01$	&	$17.5\pm0.2$		\\
1 AGeV $^{139}$La$+^{12}$C				&	$1.1$			&$1.8\pm0.1$			& $19\pm15$		& $-18\pm3$	& $0.973\pm0.008$	&	$18.3\pm0.2$		\\
1 AGeV $^{197}$Au$+^{12}$C				&	$1.3$			&$1.1\pm0.1$			& $3\pm23$		& $7\pm6$	& $1.007\pm0.007$	&	$17.7\pm0.1$		\\
1 GeV$/$c $\pi+^{197}$Au				&	$3.2$			&$0.0\pm3\times 10^{-4}$	& $151.1\pm0.7$	& $-3.8\pm0.2$	& $1.032\pm0.001$	&	$17.26\pm0.02$	\\
\hline
\end{tabular}
\end{center}
\label{results}
\end{table*}

The fragment charge yields are given by
\begin{eqnarray}
Y_{A_{\text f}, Z_{\text f}} \left( T \right) &=& \Delta t  \left< v_{\text f} \sigma _{\rm inv}\right> q_0  \exp \left( - \frac{\Delta G}{T} \right) \nonumber \\
& = & \Delta t  \left< v_{\text f} \sigma _{\rm inv}\right> \exp \left( -\frac{\Delta \mu_{\text{nfs}}}{T} \right) \nonumber \\
& \times& q_0  A_{\text f}^{-\tau} \exp \left( - \frac{a_s A_{\text f}^{\sigma}\varepsilon}{T} \right)
\label{equation}
\end{eqnarray}
a formula that depends on several quantities that are unknown, e.g. the $\Delta t$, $q_0$, etc.\ \ However, the ratio of the yield of a fragment of a given charge at a given excitation energy $Y_{A_{\text f}, Z_{\text f}} \left( T \right)$ to some reference yield of a fragment with another charge at the same excitation energy $Y_{A_{\text f}^{\prime}, Z_{\text f}^{\prime}} \left( T \right)$ cancels the all constants of proportionality and several unknown quantities.\ \ Therefore, the ratio of charge yields was fitted with the reference yield taken as the yield of fragments with the charge equal to the lower limit of the $Z_{\text f}$ fit range listed in Table~\ref{summary}.\ \ The ratio of the charge yields is given by
\begin{eqnarray}
\frac{Y_{A_{\text f}, Z_{\text f}} \left( T \right)}{Y_{A_{\text f}^{\prime}, Z_{\text f}^{\prime}} \left( T \right)} &=&
\frac{ \left< v_{\text f} \sigma _{\rm inv}\right>}{\left< v_{\text f}^{\prime} \sigma _{\rm inv}^{\prime}\right>} 
\exp \left( \frac{\Delta \mu_{\text{nfs}}^{\prime} - \Delta \mu_{\text{nfs}}}{T} \right) \nonumber \\
& \times& \exp \left( \frac{\Delta G_{\infty}^{\prime}}{T} \right)  A_{\text f}^{-\tau} \exp \left( - \frac{a A_{\text f}^{\sigma}\varepsilon}{T} \right) \nonumber \\
& = & \Theta A_{\text f}^{-\tau} \exp \left( - \frac{a_s A_{\text f}^{\sigma}\varepsilon}{T} \right)
\label{ratio}
\end{eqnarray}
where $\Theta$ absorbs all factors other than those in Fisher's model for the bulk fluid.\ \ The only free parameters in eq.~(\ref{ratio}) are $d_2$, the secondary decay coefficient, $I_0$ (and $I_1$ for the multifragmentation data) the angular momentum parameter(s) and $b_s$, the surface entropy coefficient.\ \ For most of the fragments considered here, secondary decay results in the evaporation of $\lesssim 3$ neutrons.

Previous attempts similar to the fitting of data suggested by Eq.~(\ref{ratio}) have been made.\ \ Reference \cite{hirsch-84} fit inclusive fragment yields with a form of Fisher's model that used the liquid-drop model to parameterize the energy cost of fragment formation.\ \ However, that work did not account for finite size effects, the fragment-complement Coulomb energy or the effects of angular momentum. In references \cite{elliott-02,elliott-03} the multifragmentation data presented here was fit with a version of Fisher's model that included an ad hoc Coulomb energy parameterization, but did not account for the effects of finite size or angular momentum.\ \ In reference \cite{moretto-05} some of the compound nucleus reaction data presented here was fit with Fisher's model where all the effects of finite size, Coulomb energy and angular momentum were absorbed into a chemical potential which yielded much less physical insight than the present analysis.

Table~\ref{summary} shows the fragment charge and excitation energy range over which the fits were performed.\ \ The lower limit in the fit range of $Z_{\text f}$ for the reactions $^{64}$Ni$+^{12}$C$\rightarrow^{76}$Se was set by the available data while for the other data sets it was for fragments sufficiently large to ensure that they were emitted first, or not at all.\ \ The upper limit in the fit range of $Z_{\text f}$ for the reactions $^{58}$Ni$+^{12}$C$\rightarrow^{70}$Se and $\pi+^{197}$Au is determined by the available data while for the other data sets it is determined by the largest fragment for which the fragment, complement scheme is appropriate, i.e. $Z_{\text f} < Z_{\text s} / 2$ or the largest value of $Z_{\text f}$ present in the data, whichever is smaller.\ \ The range in excitation energy is determined by starting at excitation energy values where shell effects cease and where there are one or fewer fragments in the $Z_{\text f}$ fit range.

For the compound nucleus data, there are three fit parameters for each data set: $b$, $d_2$ and $I_0$.\ \ For the multifragmentation data, there are four fit parameters for each data set: $b$, $d_2$, $I_0$ and $I_1$.\ \ On average, there are $\sim23$ points per fit parameter.

The results for $b$, $I_0$, $I_1$ and $d_2$ are listed in Table~\ref{results}.\ \ Figure~\ref{tc} shows that the results for the critical temperature of bulk nuclear matter determined from these experiments as a function of the mass of the thermal source.\ \ The multiple points shown in Fig.~\ref{tc} for the EOS and ISiS experiments are due to the differing mass and charge of the excited thermal sources; the masses and charges of the excited thermal sources are shown in Figure~\ref{remnant-az}.

Figure~\ref{angmo-plots} shows the result for the angular momentum from the fits.\ \ Also shown in that figure are other estimates of the angular momentum.\ \ One estimate is thermal in nature and is based on the angular momentum that would be imparted by the evaporation of a source beginning with $\vec{I}=0$ \cite{moretto-??}.\ \ The thermal estimate serves as a lower limit on the estimate of the angular momentum.\ \ The other estimate is based on the impulse and impact parameter of the collision; this serves as an upper limit on the estimate of the angular momentum.\ \ The results for the estimate of the angular momentum from EOS multifragmentation data lies, to within error bars, between the two estimates.

\begin{figure}
\includegraphics[width=8.7cm]{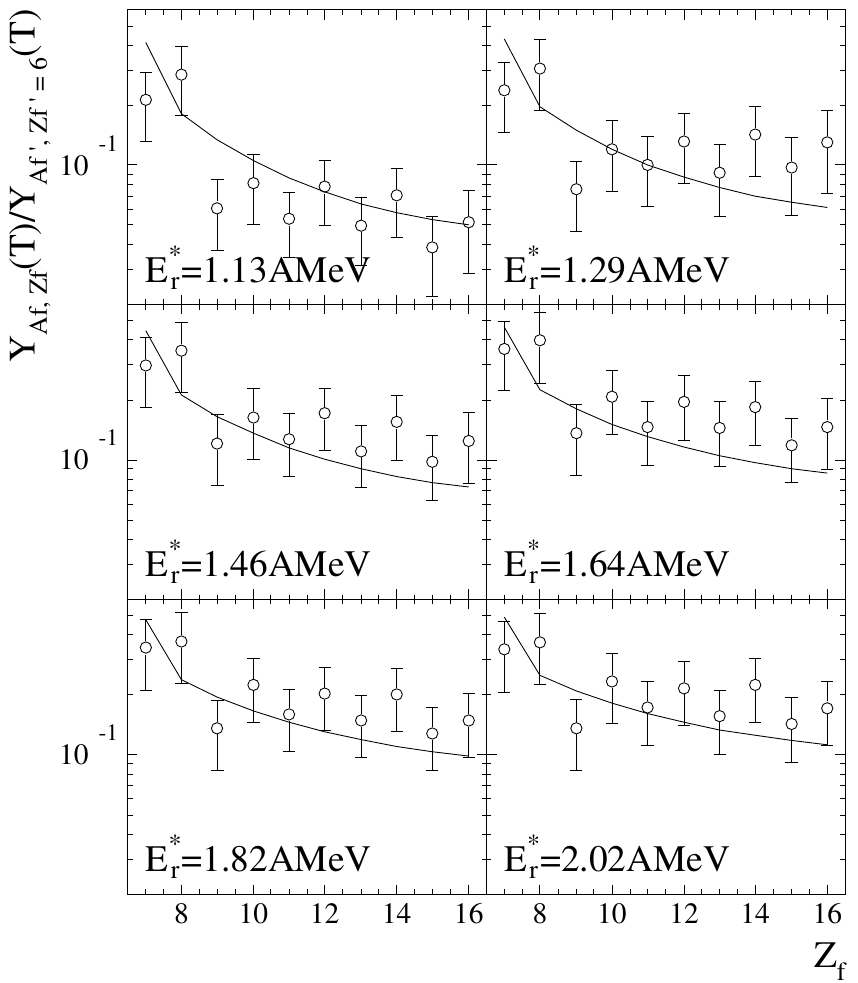}
\caption{The fragment yields from the $^{58}$Ni$+^{12}$C$\rightarrow^{70}$Se data.\ \ The curves show the fit to the data.\ \ There are 54 data points that are fit with three free parameters.}
\label{se70-yields-plots}
\end{figure}

\begin{figure}
\includegraphics[width=8.7cm]{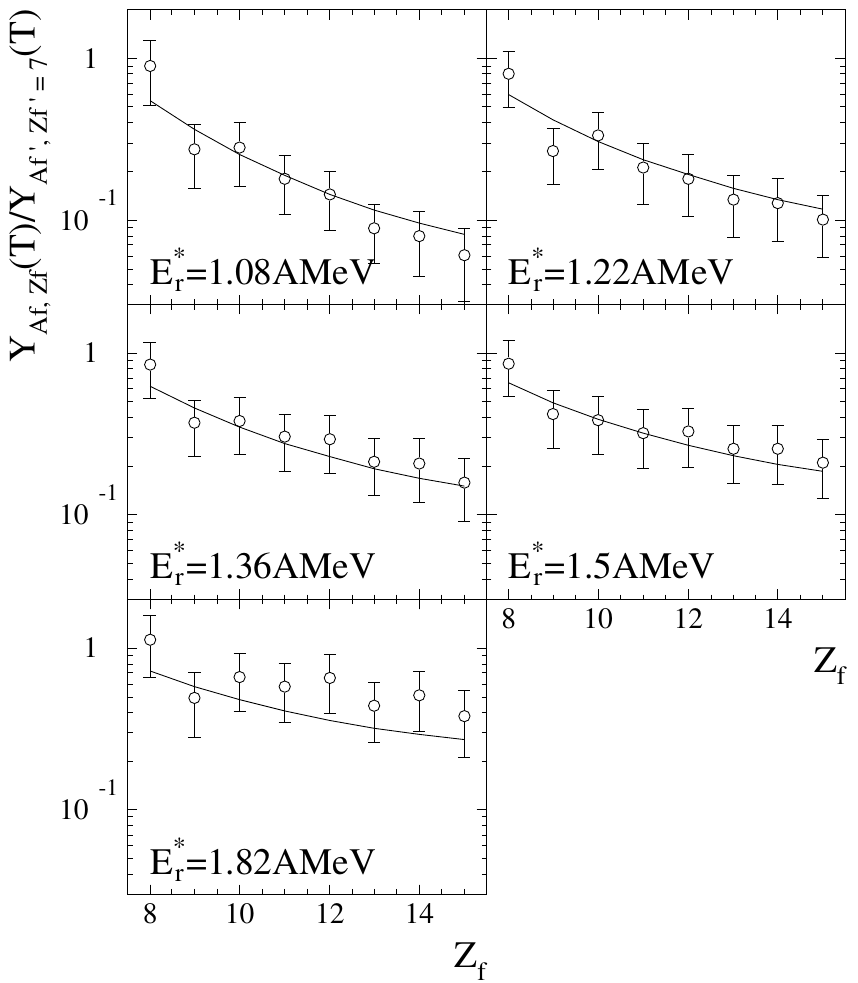}
\caption{The fragment yields from the $^{64}$Ni$+^{12}$C$\rightarrow^{76}$Se data.\ \ The curves show the fit to the data.\ \ There are 40 data points that are fit with three free parameters.}
\label{nic-yields-plots}
\end{figure}

\begin{figure}
\includegraphics[width=8.7cm]{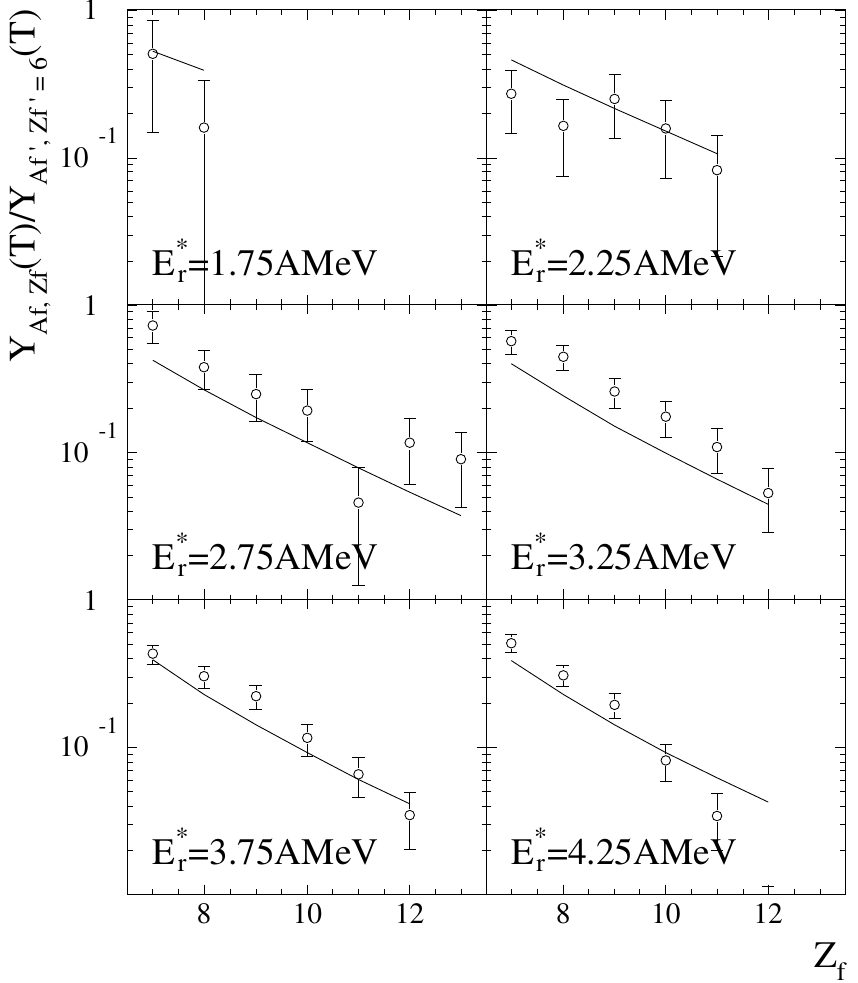}
\caption{The fragment yields from the 1 AGeV $^{84}$Kr$+^{12}$C EOS data.\ \ The curves show the fit to the data.\ \ There are 26 data points fit with four fit parameters.}
\label{krc-yields-plots}
\end{figure}

\begin{figure}
\includegraphics[width=8.7cm]{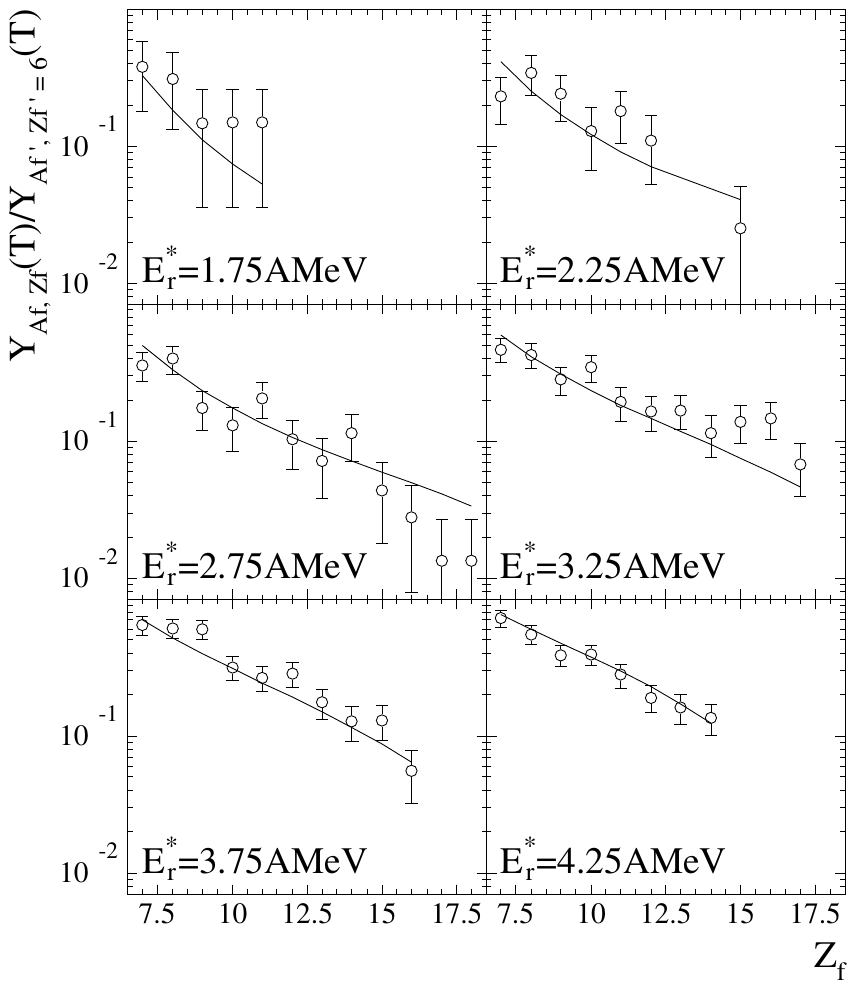}
\caption{The fragment yields from the 1 AGeV $^{139}$La$+^{12}$C EOS data.\ \ The curves show the fit to the data.\ \ There are 53 data points fit with four free parameters.}
\label{lac-yields-plots}
\end{figure}

\begin{figure}
\includegraphics[width=8.7cm]{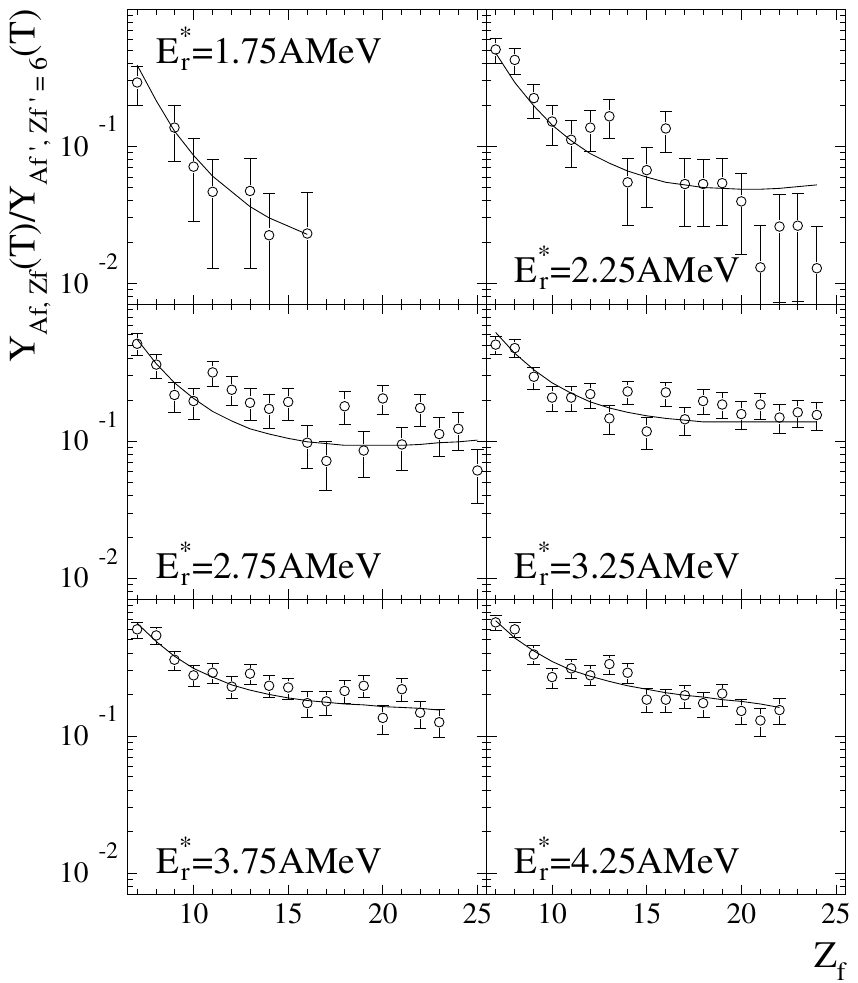}
\caption{The fragment yields from the 1 AGeV $^{179}$Au$+^{12}$C EOS data.\ \ The curves show the fit to the data.\ \ There are 96 data points fit with four free parameters.}
\label{auc-yields-plots}
\end{figure}

\begin{figure}
\includegraphics[width=8.7cm]{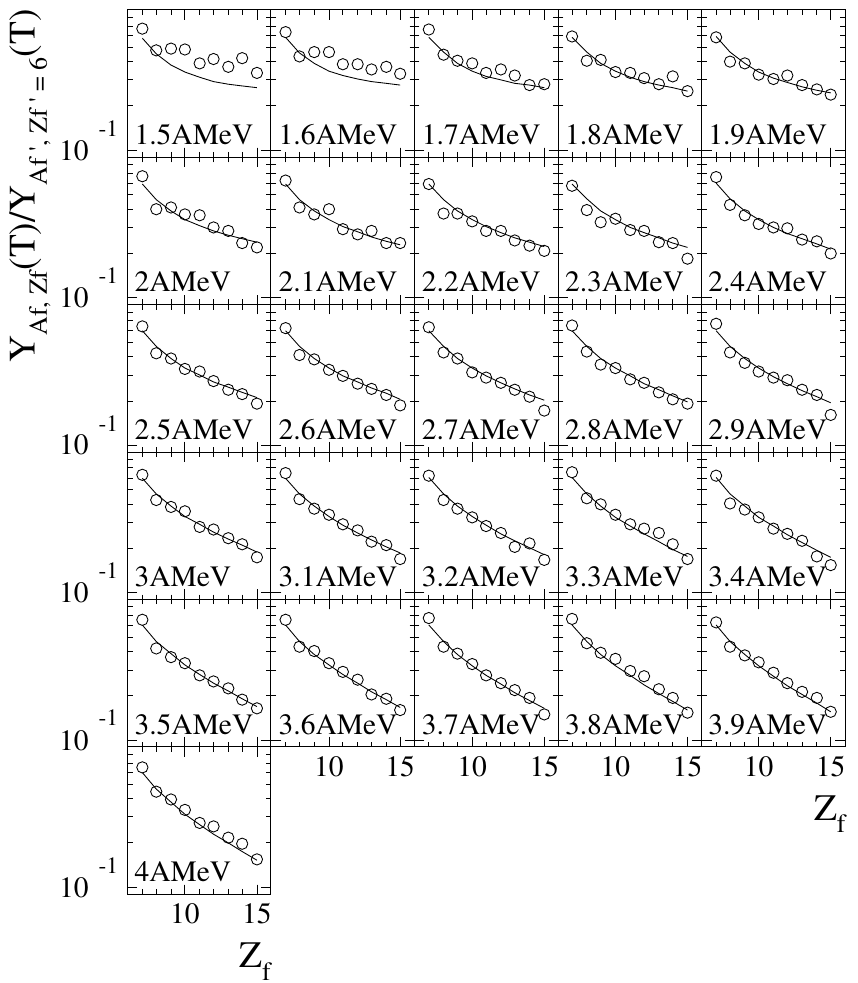}
\caption{The fragment yields from the 1 GeV$/$c $\pi+^{197}$Au ISiS data.\ \ The curves show the fit to the data.\ \ There are 234 data points fit with four free parameters.}
\label{aupi-yields-plots}
\end{figure}

\begin{figure}
\includegraphics[width=8.7cm]{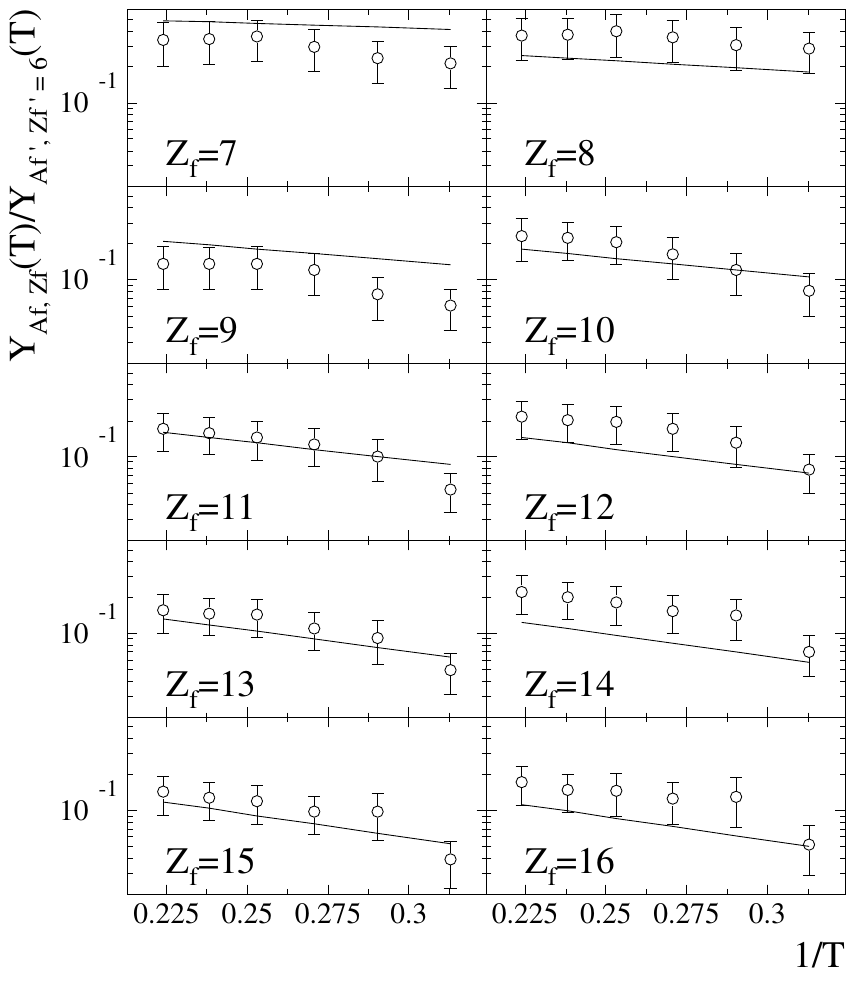}
\caption{The Arrehnius plots from the $^{58}$Ni$+^{12}$C$\rightarrow^{70}$Se data.\ \ The curves show the fit to the data.\ \ There are 54 data points fit with three free parameters.}
\label{se70-arrh-plots}
\end{figure}

\begin{figure}
\includegraphics[width=8.7cm]{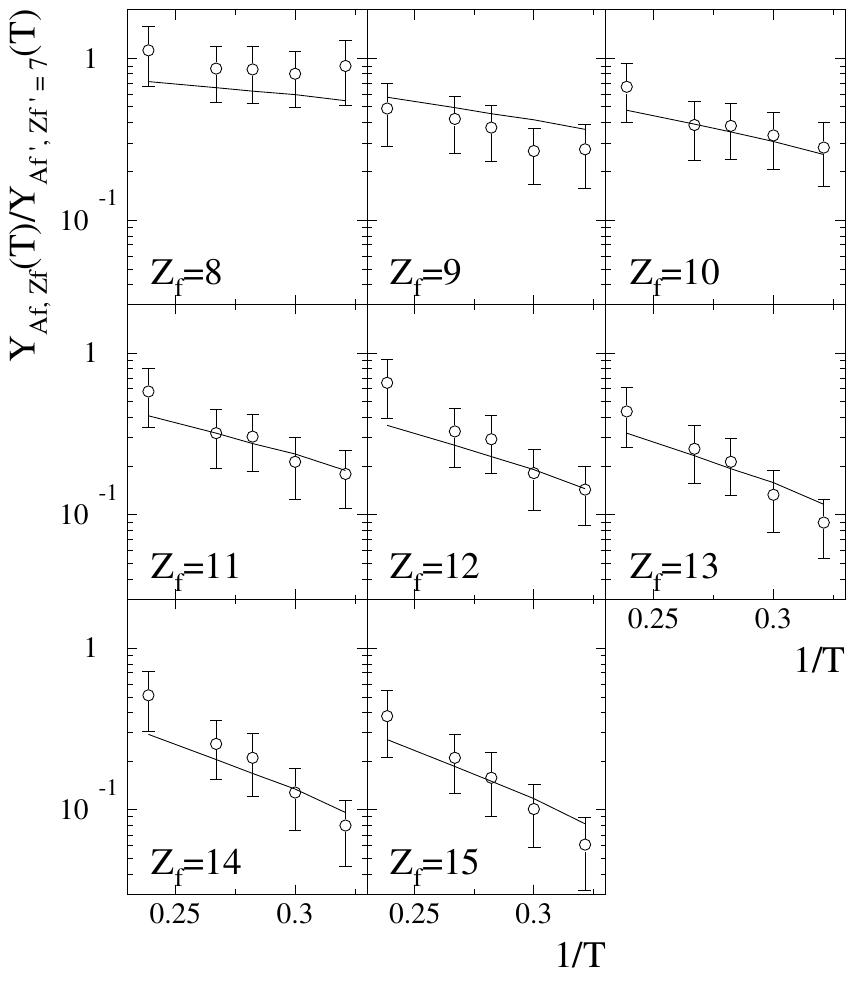}
\caption{The Arrehnius plots from the $^{64}$Ni$+^{12}$C$\rightarrow^{76}$Se data.\ \ The curves show the fit to the data.\ \ There are 40 data points fit with three free parameters.}
\label{nic-arrh-plots}
\end{figure}

\begin{figure}
\includegraphics[width=8.7cm]{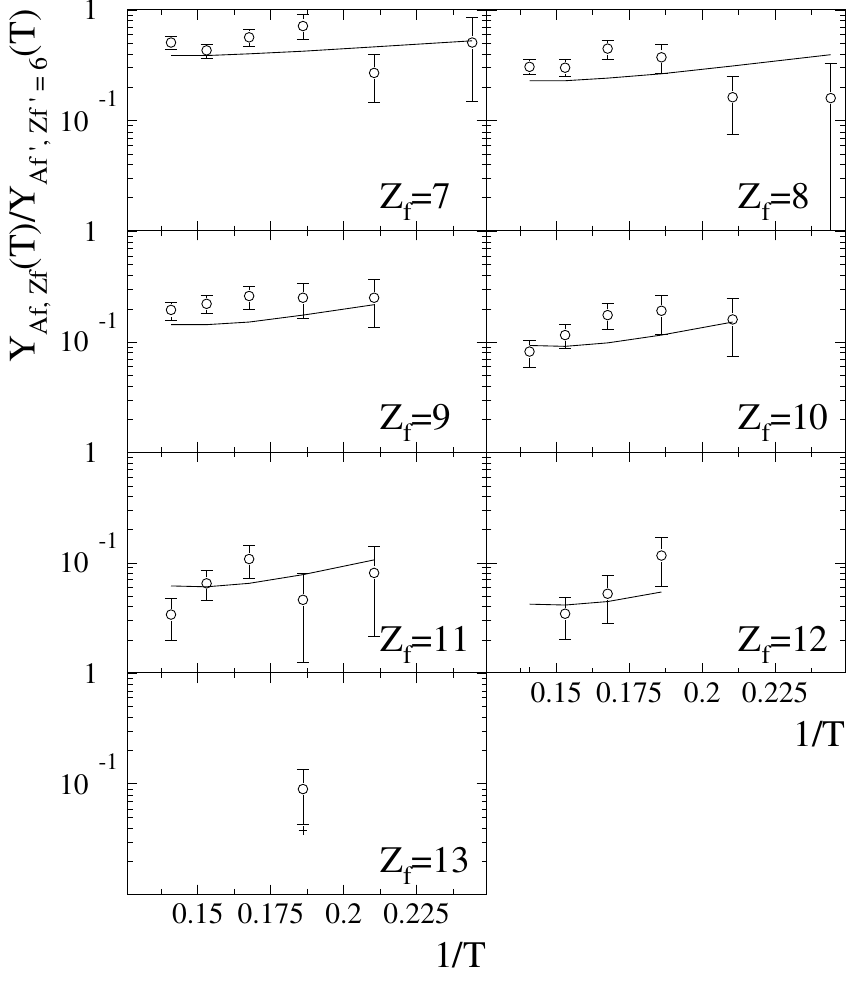}
\caption{The Arrehnius plots from the 1 AGeV $^{84}$Kr$+^{12}$C data.\ \ The curves show the fit to the data.\ \ There are 26 data points fit with four free parameters.}
\label{krc-arrh-plots}
\end{figure}

\begin{figure}
\includegraphics[width=8.7cm]{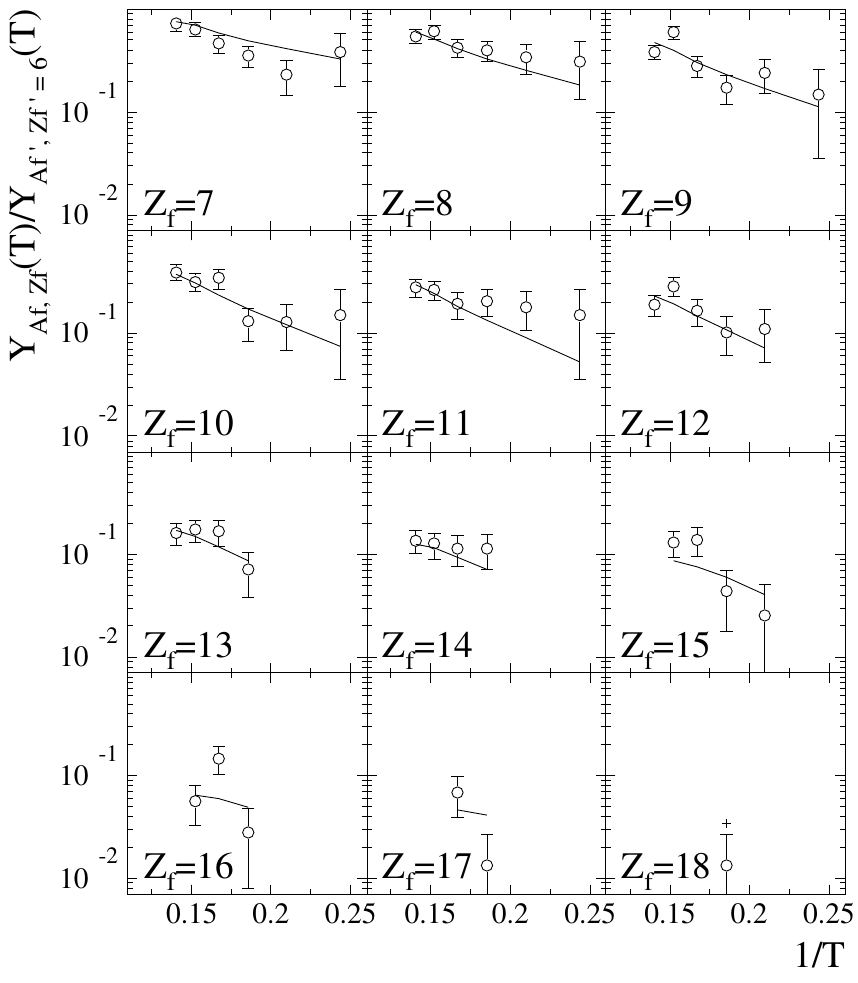}
\caption{The Arrehnius plots from the 1 AGeV $^{139}$La$+^{12}$C data.\ \ The curves show the fit to the data.\ \ There are 53 data points fit with four free parameters.}
\label{lac-arrh-plots}
\end{figure}

\begin{figure}
\includegraphics[width=8.7cm]{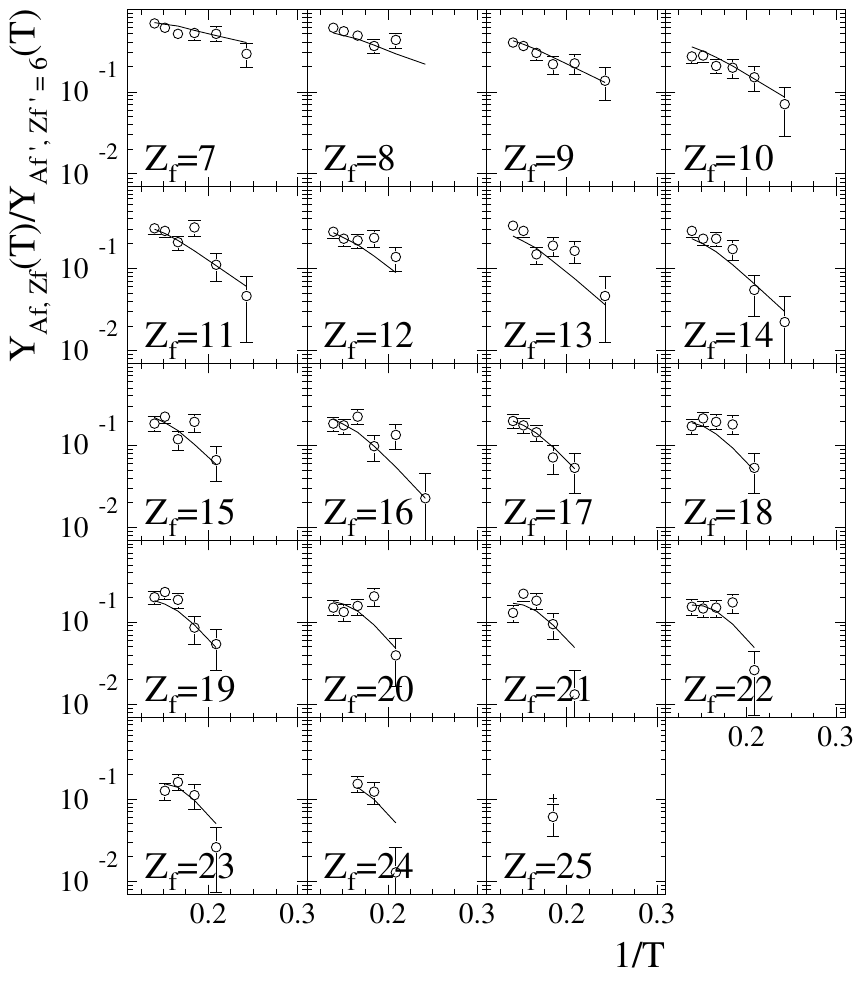}
\caption{The Arrehnius plots from the 1 AGeV $^{179}$AU$+^{12}$C data.\ \ The curves show the fit to the data.\ \ There are 96 data points fit with four free parameters.}
\label{auc-arrh-plots}
\end{figure}

\begin{figure}
\includegraphics[width=8.7cm]{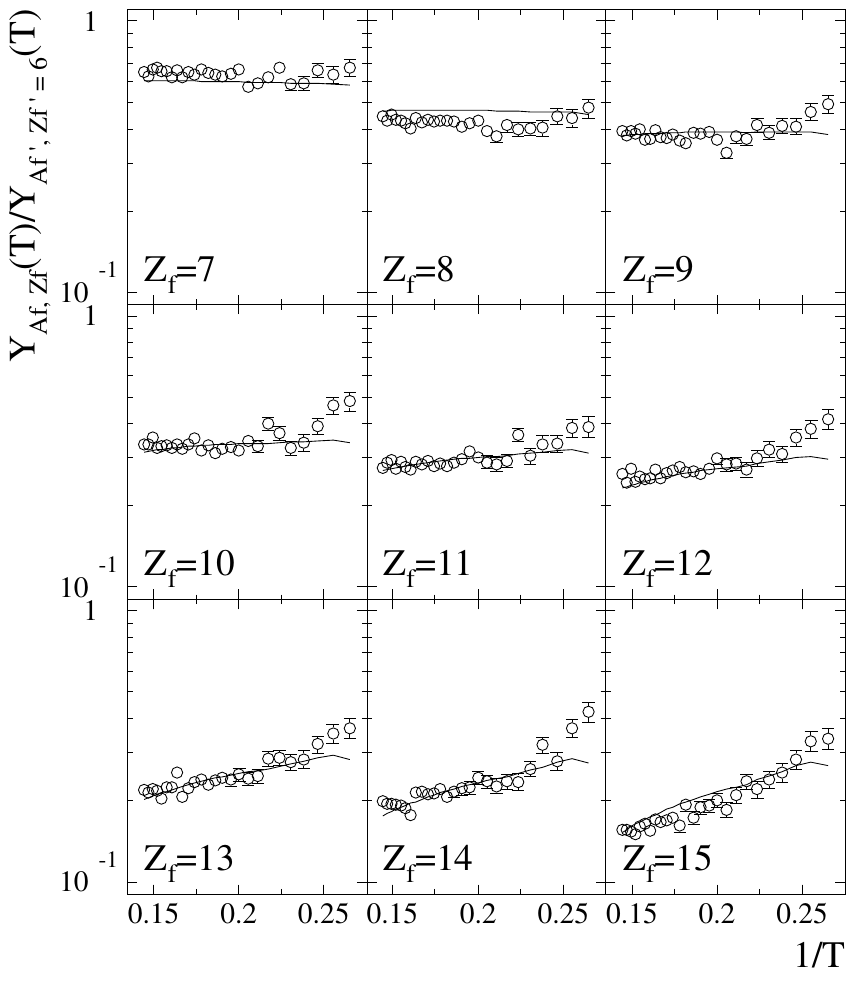}
\caption{The Arrehnius plots from the 1 GeV$/$c $\pi+^{197}$Au data.\ \ The curves show the fit to the data.\ \ There are 234 data points fit with four free parameters.}
\label{aupi-arrh-plots}
\end{figure} 

\begin{figure*}[htbp]
\includegraphics[width=18cm]{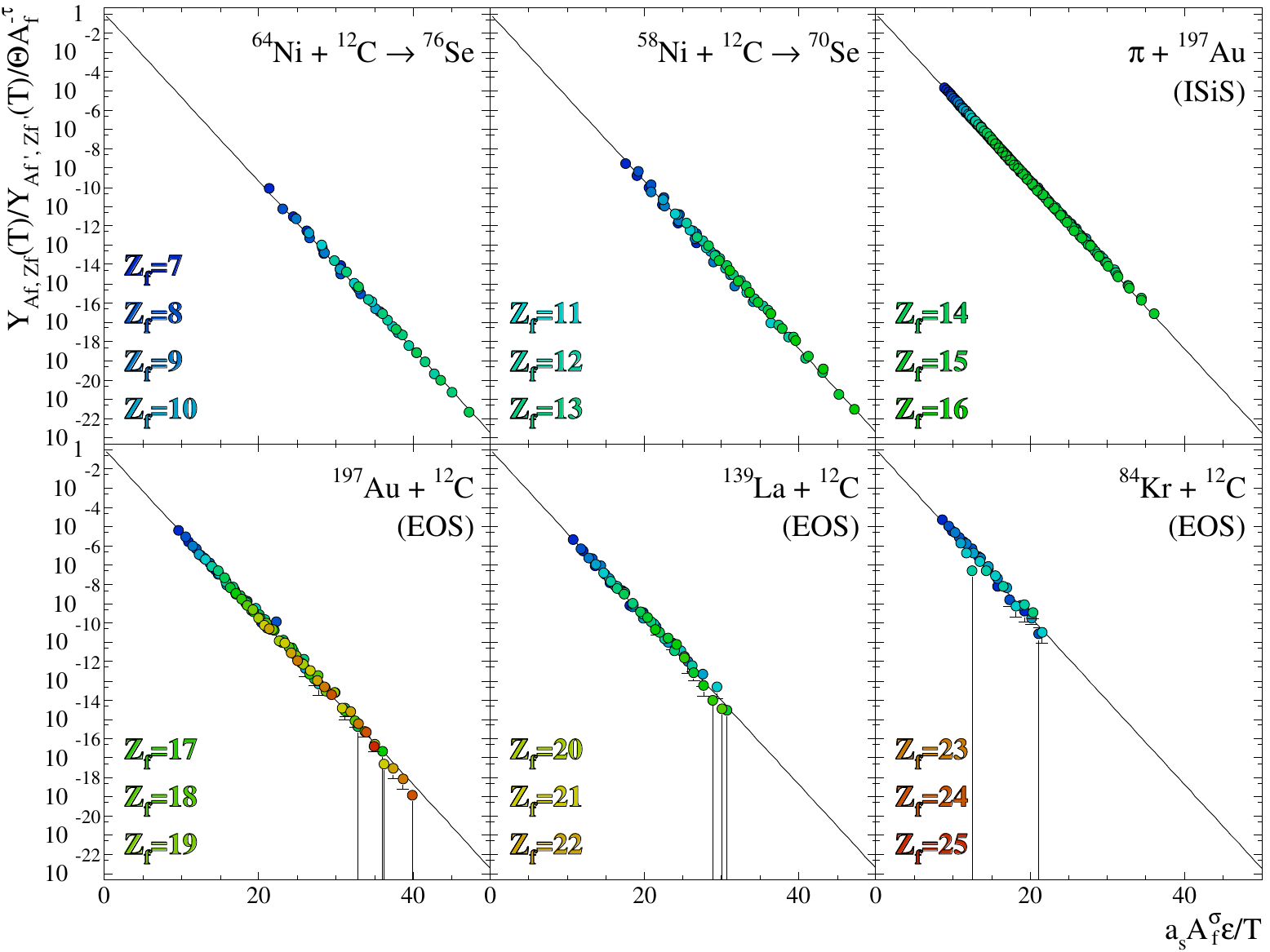}
\caption{The scaled charge yields for all six reactions.}
\label{scaling-plots}
\end{figure*}

\begin{figure}
\includegraphics[width=8.7cm]{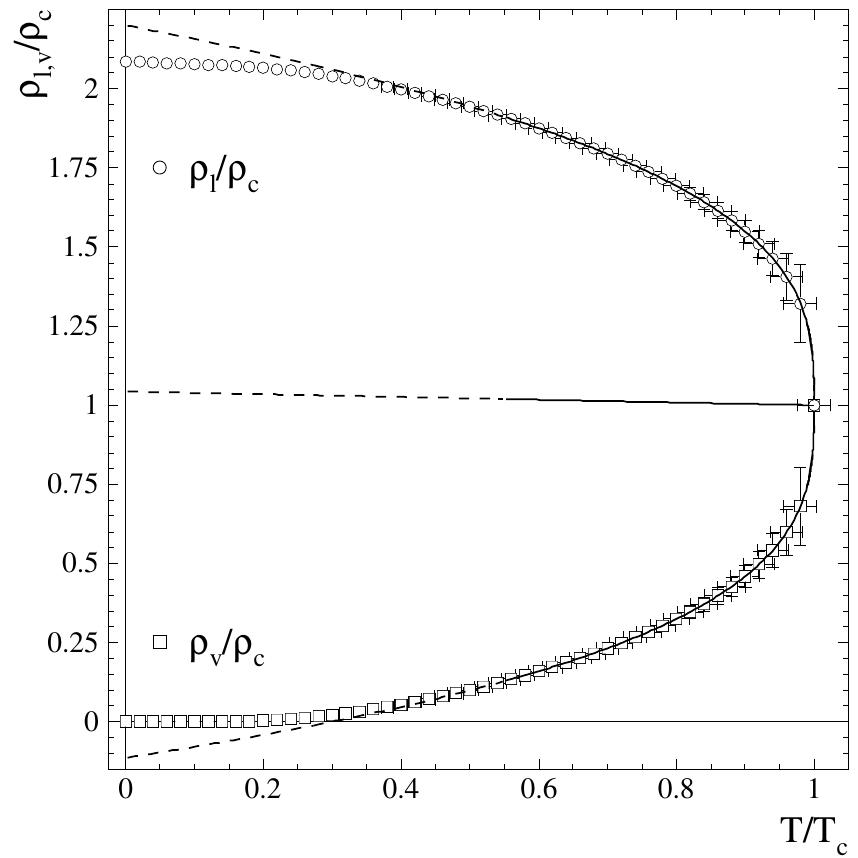}
\caption{The reduced density-reduced temperature coexistence curve for bulk nuclear matter.\ \ Empty squares show the vapor branch.\ \ Empty circles show the liquid branch.\ \ Solid curves show the results of the fit to the vapor branch.\ \ Dotted curves show the extrapolation of that fit.\ \ The dashed line shows the extrapolation of the law of rectilinear diameter.\ \ See text for details.}
\label{red-den}
\end{figure} 

The results of the surface entropy coefficient $b_s$ from all the experiments agree to within 3\%.\ \ Combining this estimate of $b_s$ with the value of $a_s$ (given below Eq.~(\ref{Ebind})) gives an estimate of the critical temperature of bulk nuclear matter as $T_c = 17.9 \pm 0.4$ MeV via Eq.~(\ref{crit-temp}).\ \ This value agrees well with theoretical predictions \cite{chomaz-04,friedman-81,glendenning-86,muller-95,jin-10}.\ \ Figure~\ref{tc} shows the value of $T_c$ as a function of the mass and charge of the thermal source.

Figures~\ref{se70-yields-plots} through \ref{aupi-yields-plots} show the fragment charge yield ratios as a function of the fragment charge.\ \ Figures~\ref{se70-arrh-plots} through \ref{aupi-arrh-plots} show Arrehnius plots in the form of the fragment charge yield ratios as a function of the inverse temperature.\ \ In all of those figures, the data is shown by the empty circles with error bars and the fits are shown with solid lines.\ \ The solid lines are segments drawn (to guide the eye) between the fit values at each $Z_{\rm f}$ or $1/T$.

Alternatively, one can combine all the results shown in figure~\ref{se70-yields-plots} through \ref{aupi-arrh-plots} by plotting $Y_{A_{\text f}, Z_{\text f}} \left( T \right)/Y_{A_{\text f}^{\prime}, Z_{\text f}^{\prime}} \left( T \right)$ divided by $\Theta A_{\text f}^{-\tau}$ as a function of ${a A^{\sigma} \varepsilon}/{T}$.\ \ This collapses the all measured fragment yield ratios for any $A_{\text f}$, $Z_{\text f}$ and $E^{*}_{\text s}$ onto a single curve.\ \ This is shown for all the data sets in Figure~\ref{scaling-plots}.\ \ Because the finite size effects, Coulomb effects, etc. have been scaled away (via dividing the yield ratios by $\Theta$), these plots show the coexistence curve of bulk nuclear matter in terms of the concentration of droplets of the bulk nuclear fluid that would comprise a saturated vapor in equilibrium with an infinite, bulk nuclear liquid.

In the plots shown in Figures~\ref{se70-yields-plots} through \ref{scaling-plots} all of the data (all fragments of all charges and all excitation energies) for a given experiment were fit simultaneously.\ \ The circles show the data points (errors are shown when they are larger than the symbols).\ \ The solid curves show the fit to the data.

\section{Constructing the phase diagram}

The fitting of the data as illustrated above gave the critical temperature of bulk nuclear matter.\ \ Once $T_c$ is determined, it is possible to determine the entire coexistence curve of bulk nuclear matter which completely maps the liquid-vapor phase diagram.

The first step is to determine the coexistence curve in reduced units:
\begin{equation}
\frac{p}{p_c}, \frac{\rho}{\rho_c}~{\rm and}~\frac{T}{T_c}.
\label{reduced}
\end{equation}
where $p$ is the pressure, $\rho$ is the density and the subscript ``{\it c}'' denotes values of the quantities at the critical point.

It is assumed that the formation of fragments exhausts all non-idealities, so that the pressure and density can be obtained by simple sums.\ \ The pressure is
\begin{eqnarray}
p & = & T \sum_A n_A(T) \nonumber \\
& = & T \sum_A q_0 A^{-\tau} \exp \left( -\frac{a_s A^{\sigma}\varepsilon}{T}\right)
\label{pressure}
\end{eqnarray}
and at the critical point
\begin{equation}
p_c = T_c \sum_A n_A(T_c) = T_c q_0 \sum_A A^{-\tau} .
\label{pressurec}
\end{equation}
The density is given by
\begin{eqnarray}
\rho & = & \sum_A A n_A(T) \nonumber \\
& = & \sum_A q_0 A^{1-\tau} \exp \left( -\frac{a_s A^{\sigma}\varepsilon}{T}\right)
\label{density}
\end{eqnarray}
and at the critical point
\begin{equation}
\rho_c = \sum_A A n_A(T_c) =  q_0 \sum_A A^{1-\tau} .
\label{densityc}
\end{equation}
Using the reduced quantities removes the unknown normalization $q_0$.\ \ All other quantities in the above sums are known.\ \ The errors associated with $T_c$, $\tau$ and $\sigma$ are propagated to generate errors on the reduced quantities.

\subsection{Reduced density}

The empty squares in Figure~\ref{red-den} shows the vapor branch of the $\rho$-$T$ phase diagram of nuclear matter, albeit in reduced form.\ \ Those points were constructed by performing the sums in equations (\ref{density}) and (\ref{densityc}).

The empty circles in Figure~\ref{red-den} show the liquid branch which was determined as follows.\ \ First, Guggenheim's universal function describing the reduced $\rho_{\text l,v}/\rho_c$-$T/T_c$ phase diagram \cite{guggenheim-45}
\begin{equation}
\frac{\rho_{\text l,v}}{\rho_c} = 1 + d_1 \varepsilon \pm d_{\beta} \varepsilon^{\beta}
\label{gugv}
\end{equation}
(where $\varepsilon$ is given by eq. (\ref{eps})) was fit to the empty squares on Figure~\ref{red-den} from $0.55T_c \le T \le T_c$ which is roughly the range over which Guggenheim's function describes dozens of fluids: from the triple point to the critical point.\ \ Here $d_1$ and $d_{\beta}$ are left as fit parameters and $\beta$ is a critical exponent and is \cite{fisher-67.1,campostrini-02}
\begin{equation}
\beta = \frac{\tau-2}{\sigma} = 0.3265 \pm 0.0001 .
\label{beta}
\end{equation}
The vapor branch is described by the eq.~(\ref{gugv}) with the minus sign and the liquid branch is described by the eq.~(\ref{gugv}) with the plus sign.

The solid curve on Figure~\ref{red-den} shows the result when the empty squares (vapor branch) were fit with eq.~(\ref{gugv}) which resulted in $d_1=0.04315\pm0.00001$ and $d_{\beta}=1.15714\pm0.00001$; the errors quoted are those resulting from the fitting procedure.\ \ These values are different from those that Guggenheim found, but that is true for other fluids as well, e.g. helium and mercury \cite{ross-96}.\ \ However, eq.~(\ref{gugv}) still describes the coexistence curve of those fluids, albeit with different values for $d_1$ and $d_{\beta}$.

For the fluids that Guggenheim examined, the liquid branch (for the range $0.55T_c \le T \le T_c$) was described by equation~(\ref{gugv}) with the sign of $d_{\beta}$ changed.\ \ The solid curve shown with $T/T_c > 1$ on Figure~\ref{red-den} shows the result.

Dashed curves on Figure~\ref{red-den} show extrapolations for $T<0.55T_c$.\ \ The extrapolation for the vapor branch shows unphysical behavior with $\rho_v/\rho_c <0$ for $T/T_c < 0.25$, thus some care must be taken when determining the $\rho_{\text l}/\rho_c$-$T/T_c$ coexistence curve at low temperatures.\ \  The $\rho_{\text v}/\rho_c$-$T/T_c$ coexistence curve at low temperatures has already been determined from equations~(\ref{density}) and (\ref{densityc}).

To determine the liquid branch of the coexistence curve for low temperatures we start with the the law of rectilinear diameter \cite{guggenheim-45} which is
\begin{equation}
\frac{\rho_l + \rho_v}{2\rho_c} = 1 + d_1 \varepsilon 
\label{law-rec-dia}
\end{equation}
We extrapolated this linear function in $\varepsilon$ from $T = 0.55T_c$ to $T = 0$.\ \ This is shown by the dashed line in Figure~\ref{red-den}.\ \  We then used that extrapolation and the values of $\rho_v/\rho_c$ computed via the sums in equations~(\ref{density}) and (\ref{densityc}) (open squares on Figure~\ref{red-den}) to solve for $\rho_l/\rho_c$ at low temperatures by ``reflecting'' them about the line defined by eq.~(\ref{law-rec-dia}).\ \ Thus
\begin{equation}
\frac{\rho_{\text l}}{\rho_c} = 2 + 2d_1 \varepsilon - \frac{\rho_v}{\rho_c} .
\label{gugl}
\end{equation}
The results are shown by empty squares on Figure~\ref{red-den}.\ \ The error bars on $\rho_l/\rho_c$ are equal to the error bars on $\rho_v/\rho_c$.

\subsection{Density}

To obtain a $\rho_{\text l,v}$-$T$ coexistence curve in a non-reduced form (e.g. temperature in units of MeV and density in units of nucleons per cubic fermi) we first multiplied the temperature axis by $T_c$.\ \ Errors on the temperature scale are then given by:
\begin{equation}
\delta T = \delta T_c \left( \frac{T}{T_c} \right).
\label{dt}
\end{equation}

\begin{figure}
\includegraphics[width=8.7cm]{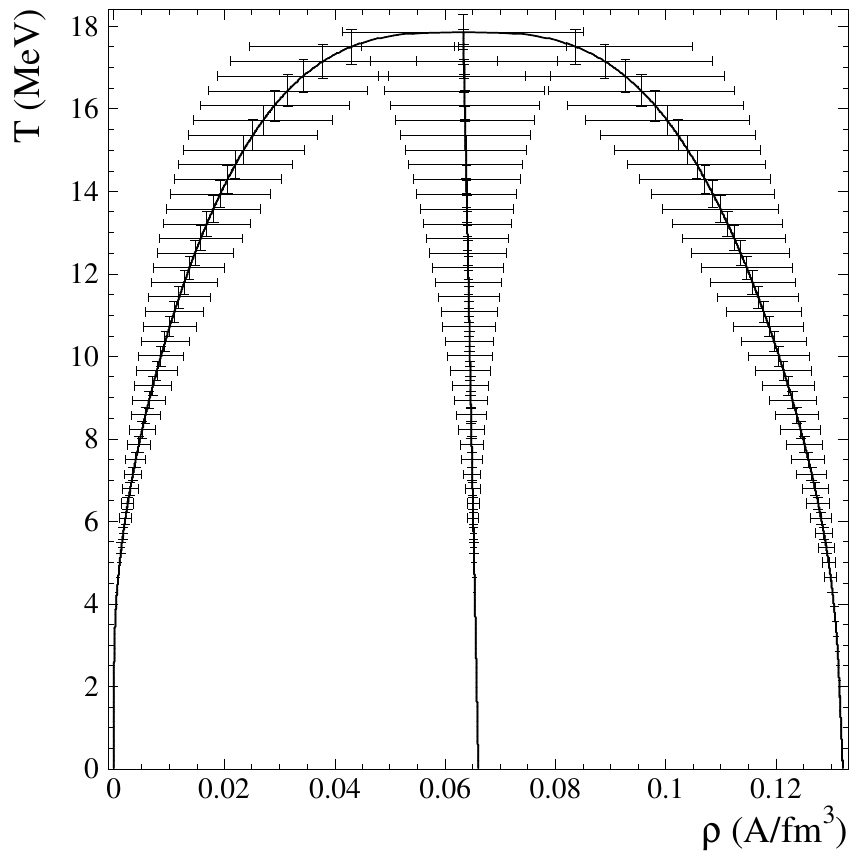}
\caption{The temperature-density coexistence curve for bulk nuclear matter.\ \ Errors are shown for selected points to give an idea of the error on the entire coexistence curve.}
\label{bulk-density}
\end{figure}

To determine the density in units of nucleons per cubic fermi we note that at $T=0$ the density of nuclear matter should be the density observed in unexcited nuclei.\ \ Using the same value of $r_0 = 1.2181~\text{fm}$ as in eq.~(\ref{Ebind}), which is within 2\% of other the leading order of other estimates \cite{myers-00}, the density of nuclear matter at $T=0$ is
\begin{equation}
\rho_l \left( T = 0 \right) = \frac{3}{4 \pi r_0^3} \approx 0.132~A/{\text {fm}}^3.
\label{liq-den}
\end{equation}
That value sets the scale on the density axis.\ \ The results are shown in Figure~\ref{bulk-density}.

The error bars shown in Figure~\ref{bulk-density} are not the same as the error bars shown in Figure~\ref{red-den}.\ \ Simply translating the errors from upper right plot of Figure~\ref{red-den} would give no estimate of the error of the critical density since, by definition there is no error associated with $\rho_c/\rho_c$.\ \ Therefore, we did the following: the error on a density value at a given temperature value $\rho ( T ) $ is
\begin{equation}
\delta \rho \left( T \right)= \frac{\left| \rho \left( T+\delta T \right) - \rho \left( T- \delta T \right)\right|}{2} .
\label{err-rho}
\end{equation}
Then we have a critical density value of $\rho_c = 0.06 \pm 0.02$ A$/$fm$^3$ which agrees well with theoretical efforts \cite{chomaz-04,glendenning-86,muller-95,jin-10}.

\subsection{Pressure}

\begin{figure}
\includegraphics[width=8.7cm]{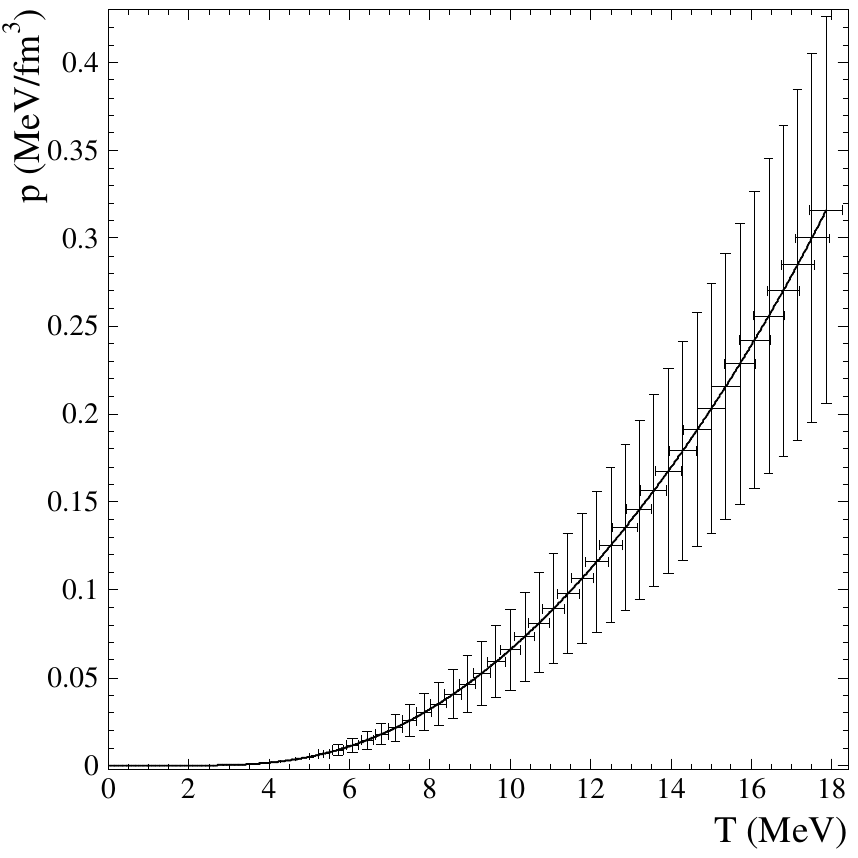}
\caption{The pressure-temperature coexistence curve for bulk nuclear matter.\ \ Errors are shown for selected points to give an idea of the error on the entire coexistence curve.}
\label{bulk-pressure}
\end{figure} 

To determine the coexistence curve for pressure as a function of temperature we again start with the reduced quantities and obtain $p/p_c$ as a function of $T/T_c$ by performing the sums in equations~(\ref{pressure}) and (\ref{pressurec}).\ \ We then determine the value of $p_c$ from the compressibility at the critical point which is defined as
\begin{equation}
Z_c = \frac{p_c}{\rho_c T_c} .
\label{compc}
\end{equation}
For fluids this is a universal quantity and is $Z_c =  0.277 \pm 0.004$.\ \ Combing equations~(\ref{pressurec}), (\ref{densityc}) and (\ref{compc}) shows that the compressibility at the critical point is just a ratio of two Riemann $\zeta$-functions \cite{kiang-70}
\begin{equation}
Z_c = \frac{\zeta \left( \tau -1 \right)}{\zeta \left( \tau \right)} = 0.276 .
\label{zetas}
\end{equation}
However, using the error on $\tau$ (given below Eq.~(\ref{rad0})) gives $Z_c = 0.28 \pm 0.01$.\ \ We use this value of $Z_c$ in combination with the values of $T_c$ and $\rho_c$ determined above to obtain a value for the pressure at the critical point of $0.3 \pm 0.1$MeV$/$fm$^3$ which agrees well with theoretical results \cite{muller-95,jin-10}.\ \ Here the error arise from the errors on $T_c$ and $\rho_c$.\ \ Now to obtain the pressure in units of MeV$/$fm$^3$ we multiply $p/p_c$ by the value of $p_c$ obtained above.\ \ The error on the pressure is given by
\begin{equation}
\delta p = \delta p_c \frac{p}{p_c} .
\label{dp}
\end{equation}
Figure~\ref{bulk-pressure} shows these results.

\section{Summary}

The aim of this paper was to extract that liquid-vapor phase diagram of infinite, uncharged, symmetric nuclear matter from the data measured in various nuclear reaction experiments using finite, charged, asymmetric nuclear matter, i. e. atomic nuclei.\ \ Because usual thermodynamical methods are obviously not accessible in this case, we concentrated on the fragment charge distributions at various excitation energies and analyzed them according to Fisher's droplet model modified to account for the finite size of the system and the nuclear nature of the fluid (e.g. isospin and Coulomb effects).

By fitting the charge yields observed in six different reactions studied in three different experiments, the critical point was found to be: $T_c = 17.9\pm0.4$ MeV, $\rho_c = 0.06\pm0.02$ A$/$fm$^3$ and $p_c = 0.3 \pm 0.1$ MeV$/$fm$^3$.\ \ Using the critical temperature and assuming the formation of fragments exhausts all non-idealities, the entire coexistence curve of bulk nuclear matter was determined from $T=0$ to the critical point.\ \ This represents the first experimental measure of the phase diagram of bulk nuclear matter and it is likely that the ideas and techniques outlined in this work would be useful in mapping other areas of the phase diagram of nuclear matter such as the phase transition between hadronic matter and the quark gluon plasma.

\section{Acknowledgments}

This work was performed by by Lawrence Berkeley National Laboratory  and was supported by the Director, Office of Energy Research, Office of High Energy and Nuclear Physics, Division of Nuclear Physics, of the U.S. Department of Energy under Contract No. DE-AC02-05CH11231.

This work also performed under the auspices of the U.S. Department of Energy by Lawrence Livermore National Laboratory under Contract DE-AC52-07NA27344.

\end{document}